\documentclass[prb,preprint, amsmath,amssymb,showpacs,superscriptaddress,floatfix]{revtex4-2}
\usepackage[breaklinks=true,colorlinks,citecolor=blue,linkcolor=blue,urlcolor=blue]{hyperref}
\usepackage[table,xcdraw]{xcolor}
\usepackage{epsfig,mathrsfs,color,latexsym,subfigure,marginnote,gensymb,}
\usepackage{graphicx}
\usepackage{xcolor}
\usepackage{booktabs}
\usepackage{adjustbox}

\begin{document}
\title{High-Throughput Screening of 2D Photocatalyst Heterostructures with Suppressed Electron–Hole Recombination for Solar Water Splitting}
\author{Shivanand Yadav}
\affiliation{Department of Physics, Bundelkhand University, Jhansi 284128, India}
%\email{yshivanand@iitk.ac.in}
\author{Jainandan Kumar Modi}
\affiliation{Department of Electrical Engineering, Indian Institute of Technology, Kanpur, Kanpur 208016, India}	
\author{Raihan Ahammed}
\affiliation{Department of Physics, Indian Institute of Technology, Kanpur, Kanpur 208016, India}
\author{B. S. Bhadoria}
\affiliation{Department of Physics, Bundelkhand University, Jhansi 284128, India}
\author{Yogesh S. Chauhan}
%\email{chauhan@iitk.ac.in}
\affiliation{Department of Electrical Engineering, Indian Institute of Technology, Kanpur, Kanpur 208016, India}	
\author{Amit Agarwal}
\email{amitag@iitk.ac.in}
\affiliation{Department of Physics, Indian Institute of Technology, Kanpur, Kanpur 208016, India}
\author{Somnath Bhowmick}
\email{bsomnath@iitk.ac.in}
\affiliation{Department of Materials Science \& Engineering, Indian Institute of Technology, Kanpur, Kanpur 208016, India}
\date{\today}

%ABSTRACT-------------------------------------------------------------
\begin{abstract}
Efficient and scalable photocatalysts for solar water splitting remain a critical challenge in renewable energy research. The work presents a high-throughput first-principles discovery of two-dimensional (2D) type-II van der Waals heterostructures (vdWHs) optimized for visible-light-driven photocatalytic water splitting. We screened 482 heterostructures constructed from 60 experimentally realizable 2D monolayers and identified 148 stable type-II vdWHs with spatially separated valence and conduction band edges, out of which 65 satisfy the thermodynamic redox conditions for water splitting over a broad pH range. Among these, the best two, MoTe$_2$/Tl$_2$O and MoSe$_2$/WSe$_2$, exhibit a high visible-light absorption coefficient exceeding 0.6$\times$10$^6$ cm$^{-1}$, resulting in a high power conversion efficiency of $\approx 2\%$. Quantum kinetic analysis of the hydrogen evolution reaction (HER) reveals nearly barrierless free energy profiles ($\Delta G_\mathrm{H}$ less than 0.1 eV) across multiple adsorption sites.  Our study further reveals that intrinsic interlayer electric fields in these vdWHs drive directional charge separation, suppressing carrier recombination. Our results establish a design framework for using type-II 2D heterostructures as tunable and experimentally accessible 2D photocatalysts for efficient hydrogen production.
\end{abstract}

\maketitle
\section{Introduction}
The global transition toward sustainable and carbon-neutral energy technologies has placed hydrogen at the forefront of clean fuel strategies. As a high-energy-density carrier that emits only water upon combustion, hydrogen presents a compelling alternative to fossil fuels~\citep{qu2013progress}. Among various production routes, photocatalytic water splitting stands out as a direct and scalable method for converting abundant solar energy into chemical energy stored in molecular hydrogen and oxygen~\citep{liu2018unique}. This light-driven process offers a transformative pathway for meeting long-term energy demands without greenhouse gas emissions~\citep{hoffmann1995environmental}. Achieving high photocatalytic efficiency, however, requires the simultaneous optimization of three tightly coupled mechanisms: strong absorption of visible light, effective spatial separation of photogenerated charge carriers to suppress recombination, and catalytic activation of redox reactions, the hydrogen evolution reaction (HER) and oxygen evolution reaction (OER), on the catalyst surface~\citep{li2020water}. To meet these conditions, the photocatalyst must possess a band gap exceeding 1.23 eV, the thermodynamic threshold for water splitting. Additionally, the conduction band minimum (CBM) must lie above the hydrogen reduction potential (-4.44 eV), while the valence band maximum (VBM) must lie below the oxygen oxidation potential (-5.67 eV), both referenced at pH $= 0$. Finally, the optical absorption edge should fall within the visible spectrum to efficiently capture the dominant portion of solar irradiance~\citep{zhuang2014computational}.

The emergence of two-dimensional (2D) materials has opened new frontiers for energy conversion and optoelectronic technologies. These atomically thin materials offer tunable electronic, optical, and catalytic properties, along with large surface-to-volume ratios and reduced dielectric screening. Beginning with graphene and functionalized graphene, known for their exceptional electronic and magnetic properties~\citep{novoselov2004electric,bhowmick2013sensory,puri2018external}, the 2D family has expanded to include hexagonal boron nitride (h-BN)~\citep{dean2010boron}, ZnO~\citep{hong2017atomic}, phosphorene~\cite{nahas2017polymorphs,priydarshi2018strain, PhysRevB.93.165413}, and various transition-metal dichalcogenides (TMDCs)~\citep{tongay2012thermally}, many of which also show promise for light absorption and catalytic activity. These materials generate charge carriers under illumination in photocatalysis, but short lifetimes and rapid electron–hole recombination often limit their efficiency. One effective strategy to mitigate these losses is to vertically stack two distinct monolayers to form a van der Waals heterostructure (vdWH). When such a heterostructure exhibits type-II band alignment, photogenerated electrons and holes naturally separate across layers, resulting in real-space charge separation and enhanced carrier lifetimes. Beyond this intrinsic advantage, vdWHs offer flexible stacking configurations that enable fine-tuning optical gaps, exciton binding energies, and catalytic performance. Recent advances in first-principles modeling have further accelerated the rational design of vdWHs by predicting stable combinations and rapidly screening their optoelectronic properties \citep{bernardi2013extraordinary,sun2017substrate,lu2014mos,roy2013graphene}. These developments have positioned 2D heterostructures as a versatile and experimentally accessible platform for next-generation photocatalysts.

Two-dimensional vdWHs can be classified into three types based on the relative alignment of the band edges in their constituent monolayers: (i) type-I (straddling gap), where both the conduction band minimum (CBM) and valence band maximum (VBM) reside in the same layer; (ii) type-II (staggered gap), where the CBM and VBM are localized in different layers; and (iii) type-III (no bandgap), where the band edges across layers overlap in energy. Each alignment offers distinct advantages for specific applications. In particular, type-II vdWHs are highly attractive for photocatalysis and photovoltaics because they allow for spontaneous real-space separation of photoexcited electrons and holes across the interface. This interlayer separation suppresses recombination, thereby extending carrier lifetimes and enhancing quantum efficiency \citep{wang2015research,rawat2019solar,mao2023first,li2021two,he2019type,wang2013visible,priydarshi2023versatility}. Numerous 2D material platforms, including transition metal dichalcogenides (TMDCs), black phosphorene, and MXenes, have demonstrated type-II alignment when paired with suitable partners \citep{kim2015band,kang2013band,gong2013band}. Several 2D materials~\citep{rahman20162d,guo2016mxene,zhan2019computational} and heterostructures, such as HfS$_2$/PtS$_2$, MoSe$_2$/WSe$_2$, and Tl$_2$O/WTe$_2$, have been shown to satisfy photocatalytic band alignment criteria while maintaining strong absorption in the visible range \citep{PhysRevMaterials.3.124002, PhysRevB.97.165306,he2020tl2o}. These case studies highlight the potential of type-II vdWHs as efficient solar energy harvesters. However, most prior efforts have focused on isolated combinations or narrow material subsets. A comprehensive, high-throughput strategy to identify the whole landscape of stable, optically active type-II vdWHs suitable for photocatalysis remains largely unexplored. This motivates the present study.

Here, we developed a high-throughput computational framework based on \textit{ab initio} density functional theory (DFT) to systematically evaluate the structural, electronic, and optical properties of van der Waals heterostructures (vdWHs) for photocatalysis [see Fig.~\ref{fig1}(a)]. Starting from a database of 60 experimentally realizable 2D monolayers \citep{mounet2018two}, we screened 482 unique bilayer combinations and identified 148 stable type-II vdWHs with spatially separated band edges. Among these, 65 heterostructures satisfy the photocatalytic band alignment criteria for water splitting over a wide pH range. We further computed their optical absorption spectra in the visible regime (1.6–3.2 eV), identifying MoTe$_2$/Tl$_2$O, MoSe$_2$/WSe$_2$, and MoTe$_2$/HfNCl as top performers, with strong sunlight absorption. These systems also exhibit nearly barrierless hydrogen evolution reaction (HER) kinetics and intrinsic interlayer fields that promote efficient charge separation. Together, our results establish a robust computational pipeline for discovering photocatalytically active 2D heterostructures and highlight a set of promising candidates for next-generation solar-to-hydrogen energy conversion technologies.

\section{Computational Methodology}
We performed first-principles calculations using density functional theory (DFT) as implemented in the Quantum Espresso package~\citep{giannozzi2009quantum}. The exchange-correlation effect was captured using Perdew-Burke-Ernzerhof (PBE) functional within the  generalized gradient approximation (GGA)~\citep{perdew1996generalized}. Since the PBE functional typically underestimates the band gap value, we employed the hybrid Heyd-Scuseria-Ernzerhof (HSE06) functional to obtain a more accurate electronic structure~\citep{ren2019using}. The kinetic energy cut-off for the plane wave basis set was set to 50 Ry based on convergence tests. A Monkhorst-Pack~\citep{pack1977special} 12$\times$12$\times$1 k-point mesh was used for the Brillouin zone integration.
%(10$\times$10$\times$1) for PBE (HSE06) calculations. 
Cell parameters and atomic positions were relaxed until the Hellmann-Feynman forces on all atoms were less than 10$^{-3}$ Ry/a.u. The total energy convergence threshold between successive iterations was set to 10$^{-4}$ Ry. The thickness of the vacuum layer was set to more than 15 {\AA} along the $z$-direction to avoid the interaction between adjacent layers. Van der Waals interactions were accounted for using Grimme’s semi-empirical DFT-D method~\citep{grimme2006semiempirical}. 

\section{Results and discussion}
\subsection{High-throughput computational screening of 2D type-II vdWHs}
\begin{figure}
    \centering{\includegraphics[width=\linewidth]{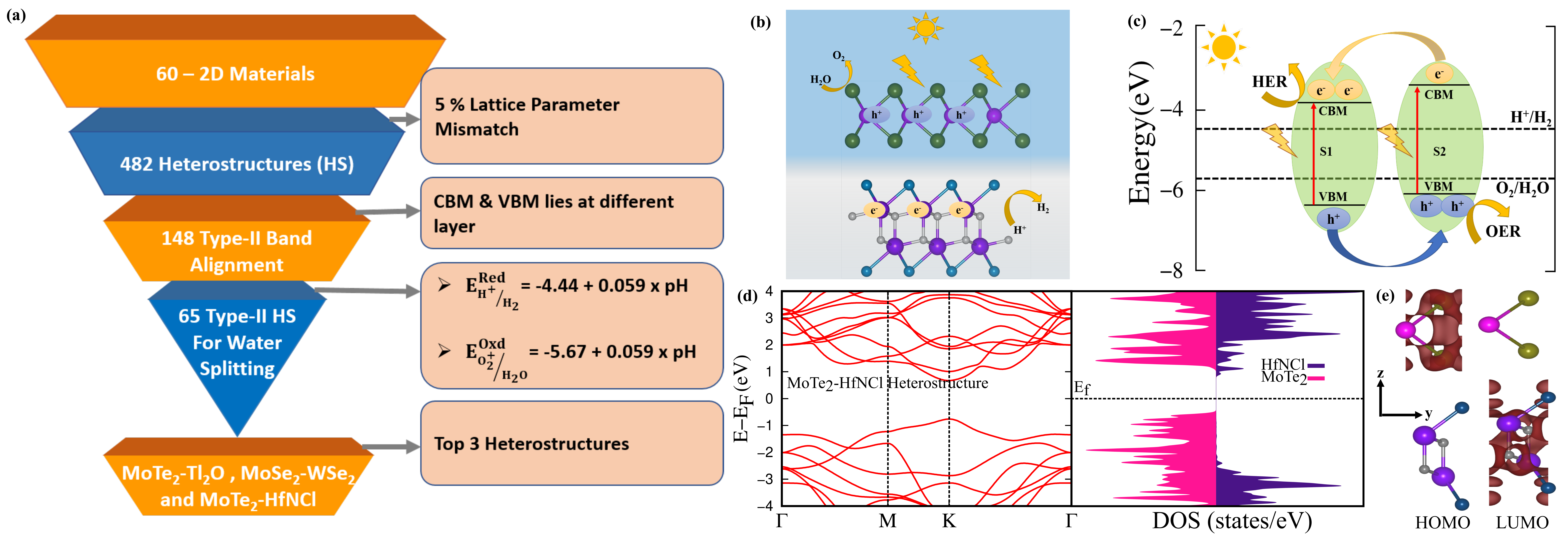}}
     \caption{(a) Schematic workflow of high-throughput first-principles calculations. Sixty-five type-II vdWHs were suitable for photocatalytic water-splitting applications out of 482 heterostructures made by combining 60 monolayers with less than 5\% lattice mismatch. (b) Schematic working principle of photocatalysis for water splitting in MoTe$_2$ (top)/HfNCl (bottom) heterostructure. The oxygen evolution reaction (OER,  2H$_2$O+4h$^+_{VB}$ = O$_2$+4H$^+$) and Hydrogen evolution reaction (HER, 2H$^+$ + 2e$^-_{CB}$=H$_2$) requires photogenerated holes, and electrons, respectively. (c) Schematic showing redox potentials at pH=0 (dashed lines), band edge alignments, and charge transfer pathways of a type-II heterostructure. (d) Band structure and density of states, (e) highest occupied molecular orbital (HOMO) and lowest unoccupied molecular orbital (LUMO) of the MoTe$_{2}$/HfNCl heterostructure, showing its type-II nature.}
    \label{fig1}
\end{figure}

Initially, we identified 60 2D monolayers from a comprehensive database~\citep{mounet2018two} and subjected them to high-throughput screening to determine the most promising heterostructures for photocatalytic water splitting. Each of these 60 monolayers has been  experimentally synthesized by exfoliation from bulk materials, ensuring their viability for practical applications. We present an  overview of the computational screening workflow used in our study in Fig.~\ref{fig1}(a). As shown in the figure, we first analyzed the crystal geometries of the heterostructures, verifying that the difference in lattice parameters between individual monolayers remained within an acceptable range of 5\%. Such minor mismatches are generally tolerable in experimental synthesis due to the inherent flexibility at the interfaces~\citep{zhang2017robust}. The lattice mismatch criterion resulted in a total of 482 bilayer heterostructures. 

Next, we identified the most stable stacking configurations for all 482 heterostructures. To thoroughly explore structural diversity, we considered six distinct stacking configurations by rotating monolayers at angles of $0^\circ$, $60^\circ$, $120^\circ$, $180^\circ$, $240^\circ$, and $300^\circ$ with respect to each other [see Figs. S1-S3 and Section S1, Supporting Information (SI) for further details]. The structure having the lowest ground-state energy was identified as the most stable stacking configuration for each heterostructure.

Subsequently, we performed DFT calculations to evaluate the electronic band structure of 482 optimized 2D heterostructures. Among them, 148 exhibited type-II band alignments, an essential criterion for efficient charge separation in different layers. The binding energy was used to assess the stability of the heterostructures. We calculated the binding energy $E_{b}$ using 
\begin{equation}
E_{b} = \frac{E_{H} - E_{m1} - E_{m2}}{A}~.
\label{eq:3}
\end{equation}%                                                        
Here, E$_{H}$ is the total energy of the heterostructure, E$_{m1}$ and E$_{m2}$ are the total energies of isolated monolayers, and $A$ is the area of the optimized heterostructure. A negative binding energy implies a stable heterostructure according to the above definition. All the heterostructures had negative binding energy, confirming their stability [see Table S1, SI for further details]. 

Finally, we found that 65 among 148 type-II heterostructures met the stringent criteria necessary for water splitting [see Eq.~\ref{eq:9} and related discussion]. These 65 heterostructures have the potential for a substantial advancement toward practical clean energy applications. Based on optical absorption and solar energy conversion efficiency [see Eq.~\ref{eq:4}, Eq.~\ref{eq:5}, Eq.~\ref{eq:6} and related discussion], the top three efficient heterostructures identified were MoTe$_{2}$/Tl$_{2}$O, MoSe$_{2}$/WSe$_{2}$, and MoTe$_{2}$/HfNCl. We also confirmed the dynamical stability of these three heterostructures [see Fig. S4, SI] via phonon dispersion calculation. The following discussion provides the details of electronic and optical properties, mainly focusing on the performance of the most efficient heterostructures. 

\subsection{Photocatalysis in type-II vdWHs}
As an illustrative example, let us choose MoTe$_2$/HfNCl, one of the top three efficient heterostructures. The overall water splitting reaction is 2H$_2$O = 2H$_2$ + O$_2$, which requires photogenerated holes and electrons. The oxidation reaction or oxygen evolution reaction (OER) is 2H$_2$O+4h$^+_{VB}$ = O$_2$+4H$^+$, and mainly takes place at the layer hosting the VBM, for example, the top MoTe$_2$ layer, Fig.~\ref{fig1}(b). On the other hand, the reduction reaction or hydrogen evolution reaction (HER) is 2H$^+$ + 2e$^-_{CB}$=H$_2$, and mainly occurs at the layer hosting the CBM, for example, the bottom HfNCl layer, Fig.~\ref{fig1}(b). 

A semiconductor must have a band gap exceeding 1.23 eV for effective photocatalytic water splitting. A schematic of the working principle, including the charge transfer pathways for water splitting via photocatalysis, is illustrated in Fig.~\ref{fig1}(c). As shown in the figure, an offset exists between the band edges of the constituent monolayers. The offset between CBM (VBM) of the constituent monolayers is termed as conduction band offset or CBO (valence band offset or VBO). The band offset leads to a charge transfer between the constituent monolayers. For example, in the case of MoTe$_2$/HfNCl heterostructure, electrons are transferred from the CBM of MoTe$_2$ to the CBM of the HfNCl monolayer, whereas holes are transported from the VBM of HfNCl to the VBM of the MoTe$_2$ monolayer. Type-II band alignment further helps to reduce electron-hole recombination, as the former (latter) is accumulated at the HfNCl (MoTe$_2$) layer, which ensures higher lifetime and more efficiency. 

Importantly, the VBM and CBM of the heterostructure should be favorably positioned for the water splitting reaction to occur. The specific criteria is specified by the following equations, which defines the redox potential for water splitting at different acidity levels or pH values, 
\begin{eqnarray}
\label{eq:9}
E^{\text{Red}}_{\text{H}^+/H_2} = -4.44 + 0.059\times\text{pH} ~, \\\nonumber
E^{\text{Oxd}}_{\text{O}_2/H_2\text{O}} = -5.67 + 0.059\times\text{pH} ~.    
\end{eqnarray}
Here, $E^{\text{Red}}$ and $E^{\text{Oxd}}$ represent the reduction and oxidation potentials, respectively, which must lie within the bandgap of the heterostructure [Fig.~\ref{fig1}(c)]. After determining the band edges from DFT calculations, we used Eq.~\ref{eq:9} to verify the suitability of the heterostructures for water splitting at different acidity levels. For example, at a pH of 0, the CBM should be positioned above the standard hydrogen reduction potential (H$^+$/H$_2$, $-$4.44 eV), while the VBM must be below the oxygen oxidation potential (O$_2$/H$_2$O, $-$5.67 eV) [see Eq.~\ref{eq:9}]. Below, we describe how one can use electronic band structure calculations to identify the heterostructures satisfying the criteria for photocatalytic water splitting.

\begin{figure}[ht]
\centering{\includegraphics[width=1.0\linewidth]{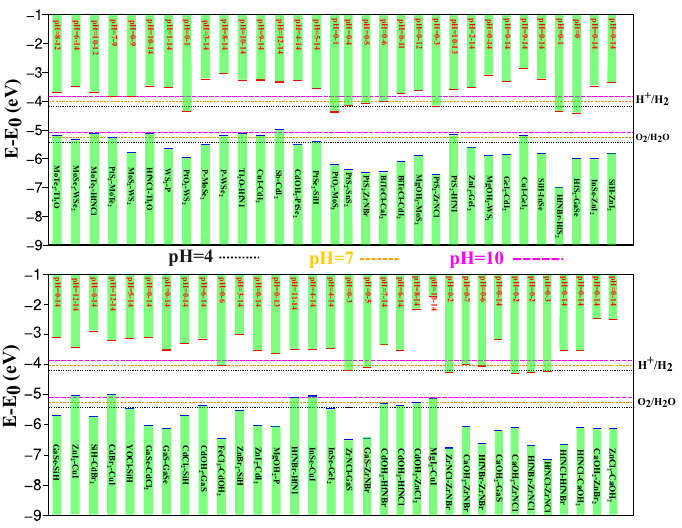}}
\caption{The band edge alignment of the 65 heterostructures is depicted in relation to the reduction potential (H${^+}$/H${_2}$) and oxidation potential (O$_2$/H$_2$O) at pH$=$4, pH$=$7, and pH$=$10, offering valuable insights into their viability for photocatalytic reactions. The y-axis shows band edges' energy ($E - E_0$), where $E_0$ is the vacuum potential energy obtained from the DFT calculations [see Eq.~\ref{eq:0}]. The pH values indicate the range in which the heterostructures satisfy the conditions for water splitting. Blue and red lines represent the VBM and CBM, respectively.}
\label{fig2}
\end{figure}

First, we calculated the band alignment with respect to the vacuum energy level from the DFT calculations using the HSE06 functional. For example, in the case of  MoTe$_2$/HfNCl heterostructure, the VBM and CBM level is located at $-5.13$ eV (MoTe$_2$) and $-3.70$ eV (HfNCl), respectively. Such a band alignment satisfies the water splitting criterion in a pH range of $\sim 10-12$, as predicted by Eq.~\ref{eq:9}. The electronic band structure and density of states (DOS) are shown in Fig.~\ref{fig1}(d), and the latter confirms a type-II band alignment, which is desirable for efficient charge separation. We further verified the type-II band alignment by plotting the highest occupied molecular orbital (HOMO) and lowest unoccupied molecular orbital (LUMO) [Fig.~\ref{fig1}(e)]. 

We performed a similar exercise for the initial list of 482 heterostructures to check their suitability for photocatalytic water splitting. First, we identified 148 type-II heterostructures by analyzing the DOS and HOMO-LUMO plot. Then, we systematically assessed band alignments and, using Eq.~\ref{eq:9}, identified 65 heterostructures satisfying the water splitting criterion at various pH levels. As shown in Fig.~\ref{fig2}, these 65 type-II heterostructures demonstrate band-edge positions compatible with water splitting requirements across various pH ranges. For example, the top two efficient heterostructures, MoTe$_{2}$/Tl$_{2}$O and WSe$_{2}$/MoSe$_{2}$ have VBM levels at $-$5.22 eV (MoTe$_{2}$) and $-$5.35 eV (WSe$_{2}$), respectively, while their CBM levels are at $-$3.73 eV (Tl$_{2}$O) and $-$3.50 eV (MoSe$_{2}$), respectively. With these band edge alignments, MoTe$_{2}$/Tl$_{2}$O and MoSe$_{2}$/WSe$_{2}$ satisfy water splitting criterion at a pH level of $\sim 8-12$ and $6-14$, respectively. The following section describes our method of ranking these 65 type-II heterostructures in terms of their optical absorption and solar energy conversion efficiency, and identifying the most promising ones for photocatalytic applications. Our study positions them as promising materials for efficient solar energy based hydrogen generation. 

\begin{figure}
    \centering{\includegraphics[width=0.95\linewidth]{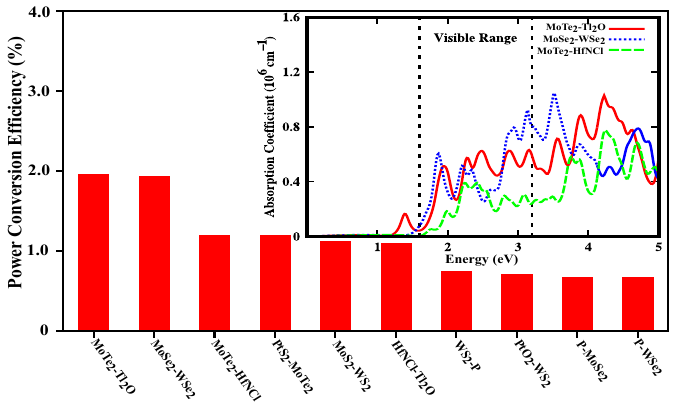}}
     \caption{The power conversion efficiency of the heterostructures. Among the 65 van der Waals heterostructures (vdWHs), 10 vdWHs are presented here, exhibiting a power conversion efficiency greater than 0.60\%, indicating their strong potential for effective energy conversion at the nanoscale. In the inset, the absorption coefficients for the three most efficient heterostructures (MoTe$_{2}$/Tl$_{2}$O, MoSe$_{2}$/WSe$_{2}$, and MoTe$_{2}$/HfNCl) are shown.} %These findings highlight the ability of these heterostructures to efficiently absorb and convert visible light, making them promising candidates for optoelectronic and photovoltaic applications.}
    \label{fig3}
\end{figure}

\subsection{Optical absorption efficiency of type-II heterostructures}
Optical absorption is one of the crucial factors for photocatalytic water splitting. The absorption coefficient can be obtained from the following formula~\citep{PhysRevB.73.045112}
\begin{equation}
	\begin{aligned}
               \mathrm{\alpha}(\mathrm{\omega})= \sqrt{2} \cdot \frac{\omega}{c}\left[\sqrt{\epsilon_1^2 (\omega) + \epsilon_2^2 (\omega)} - \epsilon_1(\omega)\right]^{1/2}~.              
\label{eq:4}
\end{aligned}
\end{equation} %
Here, $\epsilon_1(\omega)$ and $\epsilon_2(\omega)$ are the real and imaginary parts of the complex dielectric function, $\epsilon(\omega)$. c is the light velocity, and $\omega$ is the frequency. The absorbance of the heterostructures, A($\lambda$) as a function of the photon wavelength $\lambda$ can be evaluated according to the formula,
\begin{equation}
	\begin{aligned}
              A(\lambda) = 1 - e^{-\mathrm{\alpha} t} = 1 - e^{-(\frac{2 \pi}{\lambda}) \epsilon_2  t}~.
 \label{eq:5}              
\end{aligned}
\end{equation}% 
Here, $t$ is the thickness of the heterostructure and $\mathrm{\alpha}$ represents the absorption coefficient. The estimated upper limit to the converted power $P$ is based on the overlap between the solar spectrum and the absorbance, which can be obtained using the formula,
\begin{equation}
	\begin{aligned}
              P = \frac{\int_0^{\lambda_{\text{max}}} W(\lambda) A(\lambda) C(\lambda) \, d\lambda} {\int_0^{\infty} W(\lambda) \, d\lambda} ~.
	\end{aligned}
\label{eq:6}
\end{equation}%
Here, $\lambda_{\text{max}}$ is the longest wavelength that can be absorbed by the heterostructure and is determined by the band gap ($E_g$), $\lambda_{\text{max}}=\frac{hc}{E_g}$ and $W(\lambda)$ is the solar spectral irradiance at Air Mass 1.5. C($\lambda$) is the conversion factor to account for the fraction of the photon energy converted to the electron excitation energy (i.e., the thermal ionization loss), $C(\lambda)=\lambda \frac{E_g}{hc}$.

To achieve optimal efficiency for photocatalysis, materials must exhibit strong and clearly defined optical absorption within the visible spectrum. We calculated the optical absorption spectra of 65 identified heterostructures obtained from the previous section using the HSE06 functional. The absorption coefficients were determined from Eq.~\ref{eq:4} by using both the real [$\epsilon_1(\omega)$] and imaginary [$\epsilon_2(\omega)$] components of the dielectric function [see Fig. S5, SI]. Further, we calculated the solar energy conversion efficiency using Eq.~\ref{eq:5} and Eq.~\ref{eq:6}. Fig. \ref{fig3} demonstrates the top 10 heterostructures in terms of the power conversion efficiency, ranging between $\approx0.6$ to 2\%. Considering the fact that similar 2D materials and heterostructures are reported to have power conversion efficiency ranging between $\approx0.2$ to 1.5\%~\citep{1bernardi2013extraordinary, 2furchi2014photovoltaic, 3aparicio20252d, 4peng2016electronic}, these heterostructures listed in Fig.~\ref{fig3} are very promising in terms of solar energy absorption in the visible range. The inset of Fig.~\ref{fig3} presents the calculated absorption coefficients of the top 3 heterostructures, MoTe$_{2}$/Tl$_{2}$O, WSe$_{2}$/MoSe$_{2}$, and MoTe$_{2}$/HfNCl. Specifically, MoTe$_2$/HfNCl exhibits two prominent absorption peaks at approximately 2.24 eV and 2.42 eV within the visible region. Similarly, WSe$_{2}$/MoSe$_2$ displays distinct peaks at about 2.94 eV and 3.14 eV, while MoTe$_2$/Tl$_2$O shows clear absorption peaks at around 2.27 eV and 2.49 eV. Additionally, multiple absorption peaks were observed in the ultraviolet region for all studied heterostructures.  

Our calculations show that the static real dielectric constants for the MoTe$_{2}$/Tl$_{2}$O, WSe$_{2}$/MoSe$_{2}$, and MoTe$_{2}$/HfNCl are 4.92, 4.81, and 3.02, respectively. These values are considerably lower than that of silicon ($\sim$ 11.7), resulting in lower refractive indices and naturally reduced reflection losses. Consequently, these heterostructures offer improved optical efficiency with less reliance on complex anti-reflective coatings. These distinctive optical characteristics indicate that the examined heterostructures have significant potential for enhancing photocatalytic activity and solar energy harvesting efficiency. Details of the optical and electronic properties of all the top 10 heterostructures are given in Fig.~S6 and Fig.~S7, SI. The following section investigates their electronic properties in detail, further emphasizing their suitability for photocatalytic applications.

\begin{table}
\caption{Optimized lattice parameter $a$, work function ($\phi$), band edges (E$_{CBM}$ and E$_{VBM}$), band gap E$_{g}$ of isolated monolayers and heterostructures, and band offsets (CBO and VBO) of heterostructures. Band edges are calculated with reference to the vacuum potential energy $E_0$, obtained from the DFT calculations [see Eq.~\ref{eq:0}]. In case of the heterostructures, the monolayer hosting the VBM and CBM is also specified in the parentheses. GGA-PBE bandgap values are provided in the parentheses for comparison; all the other numbers are based on HSE06 calculations.}
	\label{table}
	\centering
	\adjustbox{max width=1\textwidth}{
		\begin{tabular} {c  c  c  c  c  c  c  c} 
			\hline 
			\hline 
			Materials & $a (= b)$ (\AA) & $\phi$ (eV) & E$_{CBM}$ (eV) & E$_{VBM}$ (eV) & CBO (eV) & VBO (eV) &  Bandgap (eV) \\ [1ex]
			\hline   
			\hline
			MoTe$_{2}$ &3.56 & 4.41 & -3.43 & -5.26  & -  & - &  1.83 (1.10) \\
			
			Tl$_{2}$O & 3.57 & 4.76 & -3.54 & -5.38 & - & - &    1.84 (0.96) \\
			
			HfNCl & 3.57 & 6.26 & -3.65 & -7.33 & - & - &        3.68 (2.45) \\
			
			MoSe$_{2}$ & 3.31 & 5.24 & -3.59  & -5.72 & - & - &  2.13 (1.38) \\
			
			WSe$_{2}$ & 3.31 & 4.46 & -3.34 & -5.53 & - & - &    2.19 (1.42) \\
			
			MoTe$_{2}$/Tl$_{2}$O & 3.55 & 4.66 & -3.73 (Tl$_{2}$O)  & -5.22 (MoTe$_{2}$) & 0.11 & 0.12 & 1.49 (0.96) \\
			WSe$_{2}$/MoSe$_{2}$ & 3.36 & 4.37 & -3.50 (MoSe$_{2}$) & -5.35 (WSe$_{2}$) & 0.25 & 0.19  & 1.85 (1.04) \\
			MoTe$_{2}$/HfNCl & 3.55 & 4.54 & -3.70 (HfNCl) & -5.13 (MoTe$_{2}$) & 0.22 & 2.07 & 1.43 (0.62) \\ [1ex]
			\hline 
			\hline
	\end{tabular}}
	\label{tab1}
\end{table}

\subsection{Electronic properties and band edge alignments}
\begin{figure}[h]
	\centerline{\includegraphics[width=1\linewidth]{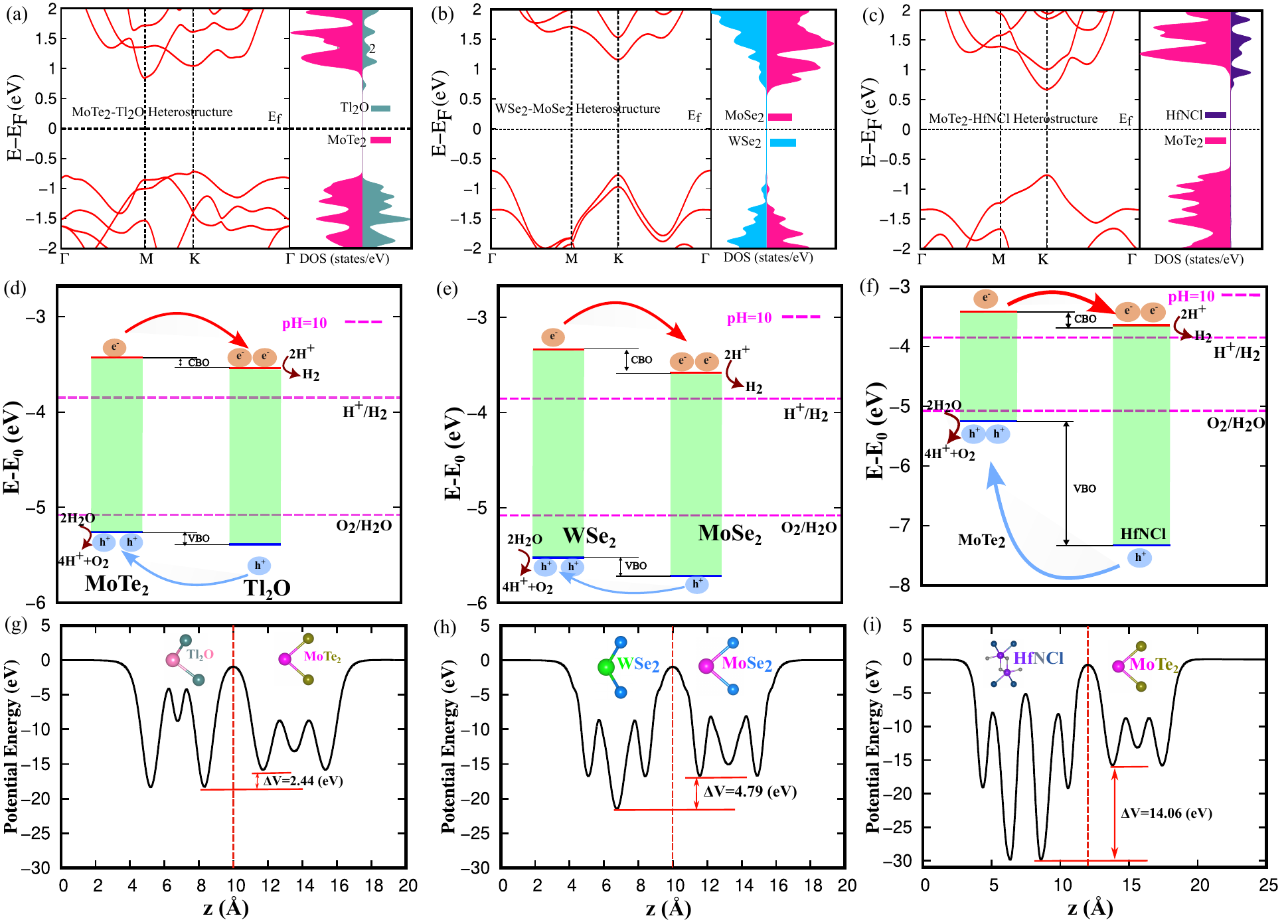}}
\caption{The electronic band structures and DOS of (a) MoTe${2}$/Tl${_2}$O, (b) MoSe${2}$/WSe${2}$, (c) MoTe$_{2}$/HfNCl heterostructures, confirming their type-II nature. Band edges and oxidation-reduction potentials at pH = 10 [marked by dotted lines, calculated using Eq.~\ref{eq:9}] of (d) MoTe${2}$/Tl${_2}$O, (e) MoSe${2}$/WSe${2}$, (f) MoTe$_{2}$/HfNCl heterostructures. The valence band offset (VBO) and conduction band offset (CBO) are indicated, emphasizing their critical role in charge transfer. The plane-averaged electrostatic potential energies [Eq.~\ref{eq:0}] of (g) MoTe$_{2}$/Tl$_{2}$O, (h) MoSe$_{2}$/WSe$_{2}$, and (i) MoTe$_{2}$/HfNCl heterostructures. The vertical dashed line shows the interface between two monolayers. The potential difference creates an electric field in the out-of-plane direction.}
\label{fig4}
\end{figure}

We analyzed the electronic band dispersion and density of states (DOS) for the constituent monolayers and heterostructures using the HSE06 functional. Essential parameters for the some constituent monolayers and 
the heterostructures, such as the lattice parameters, work function, band edges, and bandgap etc. are listed in Table~\ref{tab1}. Further details of the monolayers are provided in Fig.~S8, SI. Among the heterostructures, the electronic band structure and DOS of MoTe$_2$/Tl$_2$O, MoSe$_2$/WSe$_2$, MoTe$_2$/HfNCl, and seven other topmost efficient ones are highlighted in Fig.~\ref{fig4} (a)-(c) and Fig. S6-S7, SI, showing their type-II band alignment. MoTe$_2$/Tl$_2$O and MoSe$_2$/WSe$_2$ have indirect band gaps, while MoTe$_2$/HfNCl has a direct gap. The reduction in band gap compared to the individual monolayers arises primarily from enhanced dielectric screening due to stacking. 

We observed an offset between the band edges of the constituent monolayers [Fig.~\ref{fig4}(d)-(f)]. As reported in Table~\ref{tab1}, the conduction band offset or CBO ranges from 0.11 to 0.25 eV, causing electron transfer from MoTe$_2$ to Tl$_2$O, WSe$_2$ to MoSe$_2$, and MoTe$_2$ to HfNCl in respective heterostructures [Fig.~\ref{fig4}(d)-(f)]. As shown in the figure, due to the valence band offset or VBO, hole transfer occurs in the opposite direction, i.e.,  Tl$_2$O to MoTe$_2$, MoSe$_2$ to WSe$_2$, and HfNCl to MoTe$_2$. For the MoTe$_2$/HfNCl heterostructure, the VBO value is very large (2.07 eV), while it ranges between 0.12 and 0.19 eV for the other two heterostructures. As mentioned previously, charge separation in different monolayers minimizes electron-hole recombination. 

Work function is another indicator of charge transfer when two dissimilar materials are brought into contact. Work function $\phi$ is defined as $\phi=E_{vac}-E_F$, where the vacuum energy level is zero according to our definition. The Fermi energy is calculated with reference to the vacuum potential energy $E_0$, obtained from the DFT calculations [see Eq~\ref{eq:0} and the discussion thereafter]. Electrons are transferred from a monolayer with a lower work function to a monolayer with a higher work function. Work function values of monolayers [Table~\ref{tab1}] suggest electron flow from MoTe$_2$ to Tl$_2$O, from WSe$_2$ to MoSe$_2$, and from MoTe$_2$ to HfNCl. This is consistent with our inference based on band offset. 

\subsection{Electrostatic potential and built-in electric field}
In general, constituent monolayers of a heterostructure have different electrostatic potentials. This can be quantified by calculating the plane-averaged potential along the out-of-plane direction as,
\begin{equation}
            V(z) = \int V(x,y,z) \, dx \,dy ~,   
\label{eq:0}
\end{equation}
where $V(x,y,z)$ is obtained from the DFT calculations. Value of $V(z)$ oscillates near the monolayers and far away, becomes a constant, $E_0$, referred to as the vacuum potential energy. All the energies (like band edges, Fermi energy) are reported with reference to $E_0$.

Figures~\ref{fig4}(g)-(i) illustrate the variation of potential energy [$V(z)-E_0$] as a function of $z$, such that the vacuum energy level is set to zero for all. Figure~\ref{fig4}(g) shows that the potential energy of Tl$_2$O is deeper than that of MoTe$_2$, with a difference of $\Delta V=2.44$ eV between the two monolayers. Similarly, MoSe$_2$/WSe$_2$ [Fig.~\ref{fig4}(h)] and MoTe$_2$/HfNCl [Fig.~\ref{fig4}(i)] exhibit potential energy differences of 4.79 eV and 14.06 eV, respectively. These substantial potential energy differences give rise to an internal electric field, defined as the negative of the gradient of the electrostatic potential. In the case of MoTe$_2$/Tl$_2$O and MoTe$_2$/HfNCl heterostructures, the in-built electric field is along the positive-$z$ direction. As a result, in addition to band offset, the in-built electric field also facilitates electron transfer from  MoTe$_2$ to Tl$_2$O and from MoTe$_2$ to HfNCl. This also helps in charge separation and minimizes electron-hole recombination while  enhancing the photocatalytic efficiency.

\subsection{Gibbs free energy variation during HER}
To further probe the potential of the proposeded photocatalytic systems, we investigated the variations in reaction free energy associated with the hydrogen evolution reaction. The HER process involves two electron paths, which include a fast proton/electron transfer step and a fast hydrogen release step. These are given by, 
 \begin{equation}
	\begin{aligned}
              * + H^+ + e^- &\longrightarrow H^* ~, \\
              H^* + H^+ + e^- &\longrightarrow * + H_2(g) ~.
	\end{aligned}
\label{eq:8}
\end{equation}
Here, * represents the catalyst's active site, and $H^*$ denotes the catalyst with adsorbed $H^+$ ions. To evaluate the HER activity in MoTe${_2}$/Tl${_2}$O, MoSe${_2}$/WSe${_2}$, and MoTe$_2$/HfNCl, we calculated both the hydrogen absorption energy ($\Delta E_H$) and Gibbs free energy for hydrogen adsorption ($\Delta G_H$) [see Table~\ref{t2}], as detailed below.

First, we determined the most active and energetically favorable adsorption sites by systematically analyzing various atomic sites on the heterostructures. We employed a 3$\times$3$\times$1 supercell to model the system. Specifically, the adsorption sites considered included (a) top-Mo and top-Te atoms in the MoTe$_2$ layer and top-Tl and top-O atoms in the Tl$_2$O layer, (b) top-Mo and top-Se atoms in the MoSe$_2$ layer and top-W and top-Se atoms in the WSe$_2$ layer, and (c) top-Mo and top-Te atoms in the MoTe$_2$ layer and top-Hf, top-N, and top-Cl atoms on the HfNCl layer. These configurations are clearly illustrated in Fig.~S9, SI. 

\begin{table}[t]
\caption{Optimized adsorption distances, $\Delta E_H$, and $\Delta G_H$ for Hydrogen adsorption on different sites for MoTe$_2$/Tl2$_2$, MoSe$_2$/WSe$_2$, and MoTe$_2$/HfNCl  heterostructures. } 
	\centering
	%	\adjustbox{max width=1\textwidth}{
		\begin{tabular}{|p{0.25\linewidth}|p{0.1\linewidth}|p{0.1\linewidth}|p{0.1\linewidth}|}
			\hline
			\textbf{Hydrogen Adsorption Site} & \textbf{$D_\text{site-H}$ (\AA)} & \textbf{$\Delta E_H$ (eV)} & \textbf{$\Delta G_H$ (eV)} \\ \hline
			Mo on MoTe$_2$ & 1.71  & -0.18 & 0.06 \\ \hline
			Te on MoTe$_2$  &1.77 & -0.14  &0.10  \\ \hline
			Tl on Tl$_2$O & 1.76 & -0.16 & 0.08 \\ \hline
			O  on Tl$_2$O &0.98 & -0.17 &0.07  \\ \hline
			
		\end{tabular}

\vspace{2 mm}

	\begin{tabular}{|p{0.25\linewidth}|p{0.1\linewidth}|p{0.1\linewidth}|p{0.1\linewidth}|}
		\hline
		%\textbf{Hydrogen Adsorption Site} & \textbf{$D_\text{site-H}$ (\AA)} & \textbf{$\Delta E_H$ (eV)} & \textbf{$\Delta G_H$ (eV)} \\ \hline
		Mo on MoSe$_2$ & 1.77  & -0.15 & 0.09  \\ \hline
		Se on MoSe$_2$ & 1.53  & -0.15  & 0.09  \\ \hline
		W on WSe$_2$ & 1.53  & -0.12 & 0.12 \\ \hline
		Se on WSe$_2$ & 1.55 & -0.15 & 0.09 \\ \hline
		
	\end{tabular}
	
\vspace{2 mm}
	\begin{tabular}{|p{0.25\linewidth}|p{0.1\linewidth}|p{0.1\linewidth}|p{0.1\linewidth}|}
		\hline
		%\textbf{Hydrogen Adsorption Site} & \textbf{$D_\text{site-H}$ (\AA)} & \textbf{$\Delta E_H$ (eV)} & \textbf{$\Delta G_H$ (eV)} \\ \hline
		Mo on MoTe$_2$ & 1.7 & -0.08  & 0.16 \\ \hline
		Te on MoTe$_2$ & 1.7 & -0.08 & 0.16 \\ \hline
		Hf on HfNCl & 2.0 & -0.08 &0.16 \\ \hline
		N on HfNCl & 1.0 & -0.07  &0.17 \\ \hline
		Cl on HfNCl & 1.3 & -0.08 &0.16 \\ \hline
		
	\end{tabular}
\label{t2}
\end{table}

\begin{figure}[h]
    \centering{\includegraphics[width=0.95\linewidth]{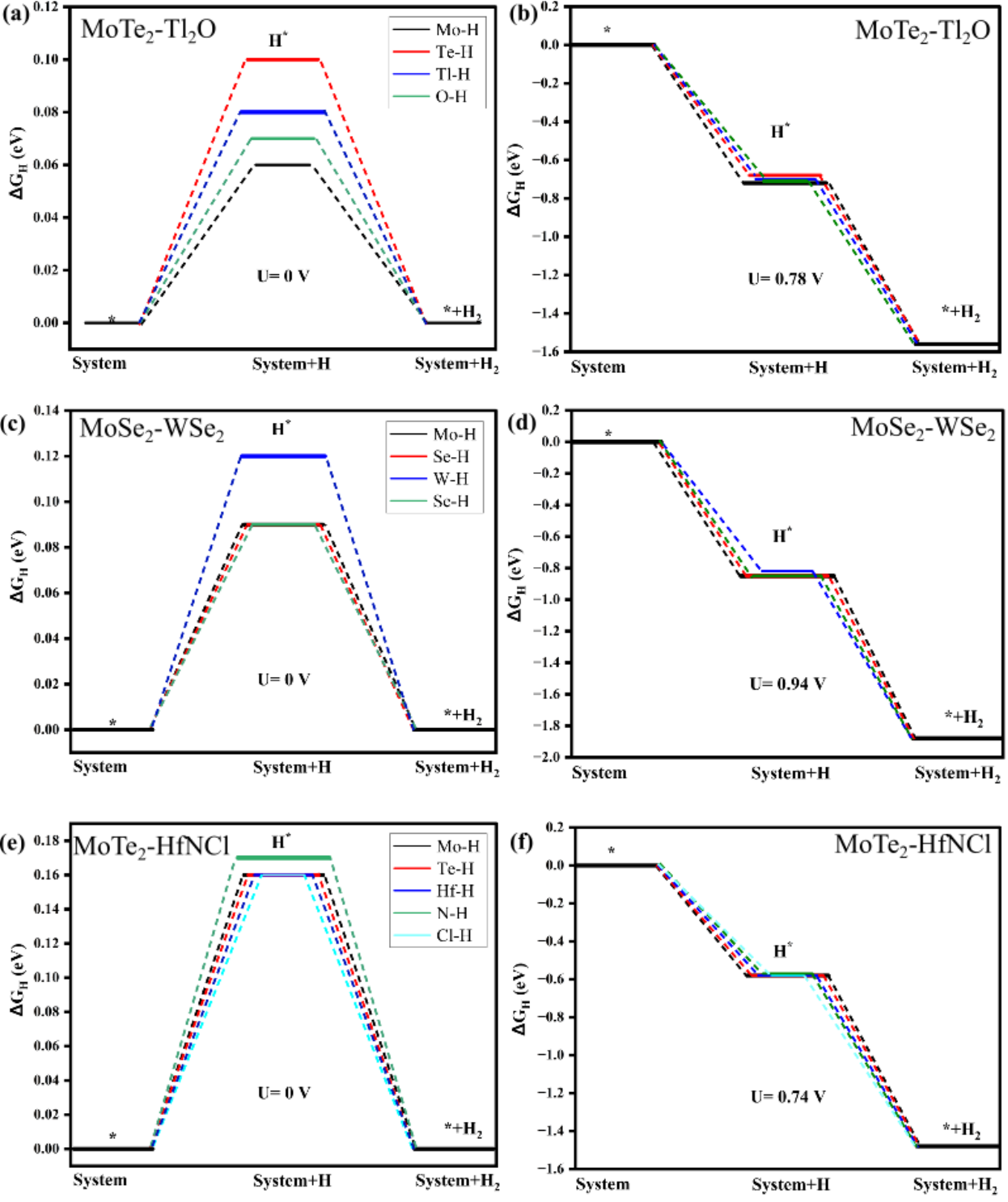}}
     \caption{Reaction paths for the hydrogen evolution reaction (HER) at different atomic sites on the MoTe$_2$ and Tl$_2$O surfaces for (a) U $=$ 0 and (b) U $=$ 0.78 V, MoSe$_2$ and WSe$_2$ surfaces for (c) U $=$ 0 and (d) U $=$ 0.94 V, and MoTe$_2$ and HfNCl surfaces for (e) U $=$ 0 and (f) U $=$ 0.74 V. * denotes the system and  H$^*$ denotes the system + H. The photogenerated electrons generate an additional potential U, resulting in a reduction in Gibbs free energy. The free energy profile ($\Delta G$), depicted in (b, d, and f), confirms that HER proceeds through a barrierless pathway.}
    \label{fig5}
\end{figure}

The Gibbs free energy ($\Delta G_H$) change for catalytic reactions using the standard hydrogen electrode (SHE) model is given by~\cite{norskov2004origin, valdes2008oxidation}, 
\begin{equation}
	\begin{aligned}
              \Delta G_H = \Delta E_H + \Delta E_{ZPE} -T\Delta S -eU ~,
	\end{aligned}
\label{eq:8a}
\end{equation}%
where $\Delta E_{ZPE}$ and $\Delta S$ denote the difference of zero-point energy and change in entropy, respectively, and $T$ is temperature.  In our calculation, we adopted the previously reported value of $\Delta E_{\text{ZPE}} - T\Delta S = 0.24$ eV \citep{nanolett.6b02052, Nørskov_2005, acsomega.1c06730, di2022universal, qu2017ultra}. In Eq.~\eqref{eq:8a}, $eU$ is the potential energy of the photogenerated electrons that drive the HER. We computed $eU$ as the difference between the conduction‐band‐edge energy and the redox potential ($E_{CBM}-E^{Red}$). The hydrogen absorption energy is calculated as
\begin{equation}
	\begin{aligned}
              \Delta E_H = E(\text{Hetrostructure + H)} - E(\text{Heterostructure}) - \frac{1}{2}E(\text{H}_2)~.
	\end{aligned}
\label{eq:9a}
\end{equation}%
Here, \( E(\text{Hetrostructure}) \) is the energy of the heterostructure, \( E(\text{Hetrostructure + H}) \) represents the energy of the heterostructure with adsorbed hydrogen, and $ E(\text{H}_2)$  is the energy of a hydrogen molecule.

For efficient HER performance, a photocatalyst ideally exhibits a $\Delta G{_H}$ value near zero, enabling easy overcoming of the energy barrier with minimal external potential. As indicated in Table~\ref{t2}, the calculated $\Delta G{_H}$ values at various adsorption sites on these heterostructures are close to zero, highlighting the effectiveness and efficiency of these heterostructures. Furthermore, Figs.~\ref{fig5} (a-f) illustrate that photogenerated electrons yield an additional potential of $U$, contributing to a decrease in the Gibbs free energy ($\Delta G < 0$). This confirms that HER proceeds through a barrierless reaction pathway, indicating the strong catalytic potential of these heterostructures.

\section{Conclusions}
We presented a comprehensive high-throughput screening of two-dimensional van der Waals heterostructures optimized for visible-light-driven photocatalytic water splitting. Out of 482 bilayer combinations screened using first-principles calculations, we identified 148 stable type-II heterostructures with spatially separated band edges, of which 65 satisfy the redox alignment criteria for efficient solar-to-hydrogen energy conversion. Three heterostructures, MoTe$_2$/Tl$_2$O, MoSe$_2$/WSe$_2$, and MoTe$_2$/HfNCl, emerged as top candidates, exhibiting strong visible-light absorption, nearly barrierless hydrogen evolution kinetics, and intrinsic interlayer electric fields that promote spatial charge separation decreasing recombination. 

Among them, MoTe$_2$/Tl$_2$O demonstrated particularly favorable properties, combining high structural stability with robust optical absorption and efficient carrier separation, yielding an estimated sunlight absorption efficiency of nearly 2\%. In addition to identifying promising material platforms for clean hydrogen generation, our work also establishes a predictive framework for designing photocatalytically active 2D heterostructures. This opens pathways for experimental realization and optimization of next-generation solar fuel technologies based on van der Waals heterostructures.

\section{Acknowledgements}
We acknowledge the National Super Computing Mission (NSM) for providing computing resources of “PARAM Sanganak” at IIT Kanpur, which is implemented by C-DAC and supported by the Ministry of Electronics and Information Technology (MeitY) and Department of Science and Technology (DST), Government of India. We also acknowledge the HPC facility provided by CC, IIT Kanpur.
\bibliography{ref}

\begin{thebibliography}{53}%
\makeatletter
\providecommand \@ifxundefined [1]{%
 \@ifx{#1\undefined}
}%
\providecommand \@ifnum [1]{%
 \ifnum #1\expandafter \@firstoftwo
 \else \expandafter \@secondoftwo
 \fi
}%
\providecommand \@ifx [1]{%
 \ifx #1\expandafter \@firstoftwo
 \else \expandafter \@secondoftwo
 \fi
}%
\providecommand \natexlab [1]{#1}%
\providecommand \enquote  [1]{``#1''}%
\providecommand \bibnamefont  [1]{#1}%
\providecommand \bibfnamefont [1]{#1}%
\providecommand \citenamefont [1]{#1}%
\providecommand \href@noop [0]{\@secondoftwo}%
\providecommand \href [0]{\begingroup \@sanitize@url \@href}%
\providecommand \@href[1]{\@@startlink{#1}\@@href}%
\providecommand \@@href[1]{\endgroup#1\@@endlink}%
\providecommand \@sanitize@url [0]{\catcode `\\12\catcode `\$12\catcode
  `\&12\catcode `\#12\catcode `\^12\catcode `\_12\catcode `\%12\relax}%
\providecommand \@@startlink[1]{}%
\providecommand \@@endlink[0]{}%
\providecommand \url  [0]{\begingroup\@sanitize@url \@url }%
\providecommand \@url [1]{\endgroup\@href {#1}{\urlprefix }}%
\providecommand \urlprefix  [0]{URL }%
\providecommand \Eprint [0]{\href }%
\providecommand \doibase [0]{https://doi.org/}%
\providecommand \selectlanguage [0]{\@gobble}%
\providecommand \bibinfo  [0]{\@secondoftwo}%
\providecommand \bibfield  [0]{\@secondoftwo}%
\providecommand \translation [1]{[#1]}%
\providecommand \BibitemOpen [0]{}%
\providecommand \bibitemStop [0]{}%
\providecommand \bibitemNoStop [0]{.\EOS\space}%
\providecommand \EOS [0]{\spacefactor3000\relax}%
\providecommand \BibitemShut  [1]{\csname bibitem#1\endcsname}%
\let\auto@bib@innerbib\@empty
%</preamble>
\bibitem [{\citenamefont {Qu}\ and\ \citenamefont
  {Duan}(2013)}]{qu2013progress}%
  \BibitemOpen
  \bibfield  {author} {\bibinfo {author} {\bibfnamefont {Y.}~\bibnamefont
  {Qu}}\ and\ \bibinfo {author} {\bibfnamefont {X.}~\bibnamefont {Duan}},\
  }\bibfield  {title} {\bibinfo {title} {Progress, challenge and perspective of
  heterogeneous photocatalysts},\ }\href
  {https://doi.org/https://doi.org/10.1039/C2CS35355E} {\bibfield  {journal}
  {\bibinfo  {journal} {Chemical Society Reviews}\ }\textbf {\bibinfo {volume}
  {42}},\ \bibinfo {pages} {2568} (\bibinfo {year} {2013})}\BibitemShut
  {NoStop}%
\bibitem [{\citenamefont {Liu}\ \emph {et~al.}(2018)\citenamefont {Liu},
  \citenamefont {Zhen}, \citenamefont {Kang}, \citenamefont {Wang},\ and\
  \citenamefont {Cheng}}]{liu2018unique}%
  \BibitemOpen
  \bibfield  {author} {\bibinfo {author} {\bibfnamefont {G.}~\bibnamefont
  {Liu}}, \bibinfo {author} {\bibfnamefont {C.}~\bibnamefont {Zhen}}, \bibinfo
  {author} {\bibfnamefont {Y.}~\bibnamefont {Kang}}, \bibinfo {author}
  {\bibfnamefont {L.}~\bibnamefont {Wang}},\ and\ \bibinfo {author}
  {\bibfnamefont {H.-M.}\ \bibnamefont {Cheng}},\ }\bibfield  {title} {\bibinfo
  {title} {Unique physicochemical properties of two-dimensional light absorbers
  facilitating photocatalysis},\ }\href
  {https://doi.org/https://doi.org/10.1039/C8CS00396C} {\bibfield  {journal}
  {\bibinfo  {journal} {Chemical Society Reviews}\ }\textbf {\bibinfo {volume}
  {47}},\ \bibinfo {pages} {6410} (\bibinfo {year} {2018})}\BibitemShut
  {NoStop}%
\bibitem [{\citenamefont {Hoffmann}\ \emph {et~al.}(1995)\citenamefont
  {Hoffmann}, \citenamefont {Martin}, \citenamefont {Choi},\ and\ \citenamefont
  {Bahnemann}}]{hoffmann1995environmental}%
  \BibitemOpen
  \bibfield  {author} {\bibinfo {author} {\bibfnamefont {M.~R.}\ \bibnamefont
  {Hoffmann}}, \bibinfo {author} {\bibfnamefont {S.~T.}\ \bibnamefont
  {Martin}}, \bibinfo {author} {\bibfnamefont {W.}~\bibnamefont {Choi}},\ and\
  \bibinfo {author} {\bibfnamefont {D.~W.}\ \bibnamefont {Bahnemann}},\
  }\bibfield  {title} {\bibinfo {title} {Environmental applications of
  semiconductor photocatalysis},\ }\href
  {https://doi.org/https://doi.org/10.1021/cr00033a004} {\bibfield  {journal}
  {\bibinfo  {journal} {Chemical reviews}\ }\textbf {\bibinfo {volume} {95}},\
  \bibinfo {pages} {69} (\bibinfo {year} {1995})}\BibitemShut {NoStop}%
\bibitem [{\citenamefont {Li}\ \emph {et~al.}(2020)\citenamefont {Li},
  \citenamefont {Zhao}, \citenamefont {Yu}, \citenamefont {Liu}, \citenamefont
  {Zhang}, \citenamefont {Liu},\ and\ \citenamefont {Zhou}}]{li2020water}%
  \BibitemOpen
  \bibfield  {author} {\bibinfo {author} {\bibfnamefont {X.}~\bibnamefont
  {Li}}, \bibinfo {author} {\bibfnamefont {L.}~\bibnamefont {Zhao}}, \bibinfo
  {author} {\bibfnamefont {J.}~\bibnamefont {Yu}}, \bibinfo {author}
  {\bibfnamefont {X.}~\bibnamefont {Liu}}, \bibinfo {author} {\bibfnamefont
  {X.}~\bibnamefont {Zhang}}, \bibinfo {author} {\bibfnamefont
  {H.}~\bibnamefont {Liu}},\ and\ \bibinfo {author} {\bibfnamefont
  {W.}~\bibnamefont {Zhou}},\ }\bibfield  {title} {\bibinfo {title} {Water
  splitting: from electrode to green energy system},\ }\href
  {https://doi.org/https://doi.org/10.1007/s40820-020-00469-3} {\bibfield
  {journal} {\bibinfo  {journal} {Nano-Micro Letters}\ }\textbf {\bibinfo
  {volume} {12}},\ \bibinfo {pages} {1} (\bibinfo {year} {2020})}\BibitemShut
  {NoStop}%
\bibitem [{\citenamefont {Zhuang}\ \emph {et~al.}(2014)\citenamefont {Zhuang},
  \citenamefont {Johannes}, \citenamefont {Blonsky},\ and\ \citenamefont
  {Hennig}}]{zhuang2014computational}%
  \BibitemOpen
  \bibfield  {author} {\bibinfo {author} {\bibfnamefont {H.~L.}\ \bibnamefont
  {Zhuang}}, \bibinfo {author} {\bibfnamefont {M.~D.}\ \bibnamefont
  {Johannes}}, \bibinfo {author} {\bibfnamefont {M.~N.}\ \bibnamefont
  {Blonsky}},\ and\ \bibinfo {author} {\bibfnamefont {R.~G.}\ \bibnamefont
  {Hennig}},\ }\bibfield  {title} {\bibinfo {title} {Computational prediction
  and characterization of single-layer crs2},\ }\href
  {https://doi.org/10.1063/1.4861659} {\bibfield  {journal} {\bibinfo
  {journal} {Applied Physics Letters}\ }\textbf {\bibinfo {volume} {104}},\
  \bibinfo {pages} {022116} (\bibinfo {year} {2014})}\BibitemShut {NoStop}%
\bibitem [{\citenamefont {Novoselov}\ \emph {et~al.}(2004)\citenamefont
  {Novoselov}, \citenamefont {Geim}, \citenamefont {Morozov}, \citenamefont
  {Jiang}, \citenamefont {Zhang}, \citenamefont {Dubonos}, \citenamefont
  {Grigorieva},\ and\ \citenamefont {Firsov}}]{novoselov2004electric}%
  \BibitemOpen
  \bibfield  {author} {\bibinfo {author} {\bibfnamefont {K.~S.}\ \bibnamefont
  {Novoselov}}, \bibinfo {author} {\bibfnamefont {A.~K.}\ \bibnamefont {Geim}},
  \bibinfo {author} {\bibfnamefont {S.~V.}\ \bibnamefont {Morozov}}, \bibinfo
  {author} {\bibfnamefont {D.-e.}\ \bibnamefont {Jiang}}, \bibinfo {author}
  {\bibfnamefont {Y.}~\bibnamefont {Zhang}}, \bibinfo {author} {\bibfnamefont
  {S.~V.}\ \bibnamefont {Dubonos}}, \bibinfo {author} {\bibfnamefont {I.~V.}\
  \bibnamefont {Grigorieva}},\ and\ \bibinfo {author} {\bibfnamefont {A.~A.}\
  \bibnamefont {Firsov}},\ }\bibfield  {title} {\bibinfo {title} {Electric
  field effect in atomically thin carbon films},\ }\href
  {https://doi.org/https://doi.org/10.1126/science.1102896} {\bibfield
  {journal} {\bibinfo  {journal} {science}\ }\textbf {\bibinfo {volume}
  {306}},\ \bibinfo {pages} {666} (\bibinfo {year} {2004})}\BibitemShut
  {NoStop}%
\bibitem [{\citenamefont {Bhowmick}\ \emph {et~al.}(2013)\citenamefont
  {Bhowmick}, \citenamefont {Medhi},\ and\ \citenamefont
  {Shenoy}}]{bhowmick2013sensory}%
  \BibitemOpen
  \bibfield  {author} {\bibinfo {author} {\bibfnamefont {S.}~\bibnamefont
  {Bhowmick}}, \bibinfo {author} {\bibfnamefont {A.}~\bibnamefont {Medhi}},\
  and\ \bibinfo {author} {\bibfnamefont {V.~B.}\ \bibnamefont {Shenoy}},\
  }\bibfield  {title} {\bibinfo {title} {Sensory-organ-like response determines
  the magnetism of zigzag-edged honeycomb nanoribbons},\ }\href
  {https://doi.org/10.1103/PhysRevB.87.085412} {\bibfield  {journal} {\bibinfo
  {journal} {Physical Review B—Condensed Matter and Materials Physics}\
  }\textbf {\bibinfo {volume} {87}},\ \bibinfo {pages} {085412} (\bibinfo
  {year} {2013})}\BibitemShut {NoStop}%
\bibitem [{\citenamefont {Puri}\ and\ \citenamefont
  {Bhowmick}(2018)}]{puri2018external}%
  \BibitemOpen
  \bibfield  {author} {\bibinfo {author} {\bibfnamefont {S.}~\bibnamefont
  {Puri}}\ and\ \bibinfo {author} {\bibfnamefont {S.}~\bibnamefont
  {Bhowmick}},\ }\bibfield  {title} {\bibinfo {title} {External-strain-induced
  semimetallic and metallic phase of chlorographene},\ }\href
  {https://doi.org/https://doi.org/10.1103/PhysRevMaterials.2.044001}
  {\bibfield  {journal} {\bibinfo  {journal} {Physical Review Materials}\
  }\textbf {\bibinfo {volume} {2}},\ \bibinfo {pages} {044001} (\bibinfo {year}
  {2018})}\BibitemShut {NoStop}%
\bibitem [{\citenamefont {Dean}\ \emph {et~al.}(2010)\citenamefont {Dean},
  \citenamefont {Young}, \citenamefont {Meric}, \citenamefont {Lee},
  \citenamefont {Wang}, \citenamefont {Sorgenfrei}, \citenamefont {Watanabe},
  \citenamefont {Taniguchi}, \citenamefont {Kim}, \citenamefont {Shepard} \emph
  {et~al.}}]{dean2010boron}%
  \BibitemOpen
  \bibfield  {author} {\bibinfo {author} {\bibfnamefont {C.~R.}\ \bibnamefont
  {Dean}}, \bibinfo {author} {\bibfnamefont {A.~F.}\ \bibnamefont {Young}},
  \bibinfo {author} {\bibfnamefont {I.}~\bibnamefont {Meric}}, \bibinfo
  {author} {\bibfnamefont {C.}~\bibnamefont {Lee}}, \bibinfo {author}
  {\bibfnamefont {L.}~\bibnamefont {Wang}}, \bibinfo {author} {\bibfnamefont
  {S.}~\bibnamefont {Sorgenfrei}}, \bibinfo {author} {\bibfnamefont
  {K.}~\bibnamefont {Watanabe}}, \bibinfo {author} {\bibfnamefont
  {T.}~\bibnamefont {Taniguchi}}, \bibinfo {author} {\bibfnamefont
  {P.}~\bibnamefont {Kim}}, \bibinfo {author} {\bibfnamefont {K.~L.}\
  \bibnamefont {Shepard}}, \emph {et~al.},\ }\bibfield  {title} {\bibinfo
  {title} {Boron nitride substrates for high-quality graphene electronics},\
  }\href {https://doi.org/10.1038/NNANO.2010.172} {\bibfield  {journal}
  {\bibinfo  {journal} {Nature nanotechnology}\ }\textbf {\bibinfo {volume}
  {5}},\ \bibinfo {pages} {722} (\bibinfo {year} {2010})}\BibitemShut {NoStop}%
\bibitem [{\citenamefont {Hong}\ \emph {et~al.}(2017)\citenamefont {Hong},
  \citenamefont {Jo}, \citenamefont {Hwang}, \citenamefont {Lee}, \citenamefont
  {Kim}, \citenamefont {Son}, \citenamefont {Kim}, \citenamefont {Jin},
  \citenamefont {Jun}, \citenamefont {Erni} \emph {et~al.}}]{hong2017atomic}%
  \BibitemOpen
  \bibfield  {author} {\bibinfo {author} {\bibfnamefont {H.-K.}\ \bibnamefont
  {Hong}}, \bibinfo {author} {\bibfnamefont {J.}~\bibnamefont {Jo}}, \bibinfo
  {author} {\bibfnamefont {D.}~\bibnamefont {Hwang}}, \bibinfo {author}
  {\bibfnamefont {J.}~\bibnamefont {Lee}}, \bibinfo {author} {\bibfnamefont
  {N.~Y.}\ \bibnamefont {Kim}}, \bibinfo {author} {\bibfnamefont
  {S.}~\bibnamefont {Son}}, \bibinfo {author} {\bibfnamefont {J.~H.}\
  \bibnamefont {Kim}}, \bibinfo {author} {\bibfnamefont {M.-J.}\ \bibnamefont
  {Jin}}, \bibinfo {author} {\bibfnamefont {Y.~C.}\ \bibnamefont {Jun}},
  \bibinfo {author} {\bibfnamefont {R.}~\bibnamefont {Erni}}, \emph {et~al.},\
  }\bibfield  {title} {\bibinfo {title} {Atomic scale study on growth and
  heteroepitaxy of zno monolayer on graphene},\ }\href
  {https://doi.org/10.1021/acs.nanolett.6b03621} {\bibfield  {journal}
  {\bibinfo  {journal} {Nano letters}\ }\textbf {\bibinfo {volume} {17}},\
  \bibinfo {pages} {120} (\bibinfo {year} {2017})}\BibitemShut {NoStop}%
\bibitem [{\citenamefont {Nahas}\ \emph {et~al.}(2017)\citenamefont {Nahas},
  \citenamefont {Bajaj},\ and\ \citenamefont {Bhowmick}}]{nahas2017polymorphs}%
  \BibitemOpen
  \bibfield  {author} {\bibinfo {author} {\bibfnamefont {S.}~\bibnamefont
  {Nahas}}, \bibinfo {author} {\bibfnamefont {A.}~\bibnamefont {Bajaj}},\ and\
  \bibinfo {author} {\bibfnamefont {S.}~\bibnamefont {Bhowmick}},\ }\bibfield
  {title} {\bibinfo {title} {Polymorphs of two dimensional phosphorus and
  arsenic: insight from an evolutionary search},\ }\href
  {https://doi.org/10.1039/c6cp08807d} {\bibfield  {journal} {\bibinfo
  {journal} {Physical Chemistry Chemical Physics}\ }\textbf {\bibinfo {volume}
  {19}},\ \bibinfo {pages} {11282} (\bibinfo {year} {2017})}\BibitemShut
  {NoStop}%
\bibitem [{\citenamefont {Priydarshi}\ \emph {et~al.}(2018)\citenamefont
  {Priydarshi}, \citenamefont {Chauhan}, \citenamefont {Bhowmick},\ and\
  \citenamefont {Agarwal}}]{priydarshi2018strain}%
  \BibitemOpen
  \bibfield  {author} {\bibinfo {author} {\bibfnamefont {A.}~\bibnamefont
  {Priydarshi}}, \bibinfo {author} {\bibfnamefont {Y.~S.}\ \bibnamefont
  {Chauhan}}, \bibinfo {author} {\bibfnamefont {S.}~\bibnamefont {Bhowmick}},\
  and\ \bibinfo {author} {\bibfnamefont {A.}~\bibnamefont {Agarwal}},\
  }\bibfield  {title} {\bibinfo {title} {Strain-tunable charge carrier mobility
  of atomically thin phosphorus allotropes},\ }\href
  {https://doi.org/10.1103/PhysRevB.97.115434} {\bibfield  {journal} {\bibinfo
  {journal} {Physical Review B}\ }\textbf {\bibinfo {volume} {97}},\ \bibinfo
  {pages} {115434} (\bibinfo {year} {2018})}\BibitemShut {NoStop}%
\bibitem [{\citenamefont {Nahas}\ \emph {et~al.}(2016)\citenamefont {Nahas},
  \citenamefont {Ghosh}, \citenamefont {Bhowmick},\ and\ \citenamefont
  {Agarwal}}]{PhysRevB.93.165413}%
  \BibitemOpen
  \bibfield  {author} {\bibinfo {author} {\bibfnamefont {S.}~\bibnamefont
  {Nahas}}, \bibinfo {author} {\bibfnamefont {B.}~\bibnamefont {Ghosh}},
  \bibinfo {author} {\bibfnamefont {S.}~\bibnamefont {Bhowmick}},\ and\
  \bibinfo {author} {\bibfnamefont {A.}~\bibnamefont {Agarwal}},\ }\bibfield
  {title} {\bibinfo {title} {First-principles cluster expansion study of
  functionalization of black phosphorene via fluorination and oxidation},\
  }\href {https://doi.org/10.1103/PhysRevB.93.165413} {\bibfield  {journal}
  {\bibinfo  {journal} {Phys. Rev. B}\ }\textbf {\bibinfo {volume} {93}},\
  \bibinfo {pages} {165413} (\bibinfo {year} {2016})}\BibitemShut {NoStop}%
\bibitem [{\citenamefont {Tongay}\ \emph {et~al.}(2012)\citenamefont {Tongay},
  \citenamefont {Zhou}, \citenamefont {Ataca}, \citenamefont {Lo},
  \citenamefont {Matthews}, \citenamefont {Li}, \citenamefont {Grossman},\ and\
  \citenamefont {Wu}}]{tongay2012thermally}%
  \BibitemOpen
  \bibfield  {author} {\bibinfo {author} {\bibfnamefont {S.}~\bibnamefont
  {Tongay}}, \bibinfo {author} {\bibfnamefont {J.}~\bibnamefont {Zhou}},
  \bibinfo {author} {\bibfnamefont {C.}~\bibnamefont {Ataca}}, \bibinfo
  {author} {\bibfnamefont {K.}~\bibnamefont {Lo}}, \bibinfo {author}
  {\bibfnamefont {T.~S.}\ \bibnamefont {Matthews}}, \bibinfo {author}
  {\bibfnamefont {J.}~\bibnamefont {Li}}, \bibinfo {author} {\bibfnamefont
  {J.~C.}\ \bibnamefont {Grossman}},\ and\ \bibinfo {author} {\bibfnamefont
  {J.}~\bibnamefont {Wu}},\ }\bibfield  {title} {\bibinfo {title} {Thermally
  driven crossover from indirect toward direct bandgap in 2d semiconductors:
  Mose2 versus mos2},\ }\href {https://doi.org/dx.doi.org/10.1021/nl302584w}
  {\bibfield  {journal} {\bibinfo  {journal} {Nano letters}\ }\textbf {\bibinfo
  {volume} {12}},\ \bibinfo {pages} {5576} (\bibinfo {year}
  {2012})}\BibitemShut {NoStop}%
\bibitem [{\citenamefont {Bernardi}\ \emph
  {et~al.}(2013{\natexlab{a}})\citenamefont {Bernardi}, \citenamefont
  {Palummo},\ and\ \citenamefont {Grossman}}]{bernardi2013extraordinary}%
  \BibitemOpen
  \bibfield  {author} {\bibinfo {author} {\bibfnamefont {M.}~\bibnamefont
  {Bernardi}}, \bibinfo {author} {\bibfnamefont {M.}~\bibnamefont {Palummo}},\
  and\ \bibinfo {author} {\bibfnamefont {J.~C.}\ \bibnamefont {Grossman}},\
  }\bibfield  {title} {\bibinfo {title} {Extraordinary sunlight absorption and
  one nanometer thick photovoltaics using two-dimensional monolayer
  materials},\ }\href {https://doi.org/10.1021/nl401544y} {\bibfield  {journal}
  {\bibinfo  {journal} {Nano letters}\ }\textbf {\bibinfo {volume} {13}},\
  \bibinfo {pages} {3664} (\bibinfo {year} {2013}{\natexlab{a}})}\BibitemShut
  {NoStop}%
\bibitem [{\citenamefont {Sun}\ \emph {et~al.}(2017)\citenamefont {Sun},
  \citenamefont {Wang},\ and\ \citenamefont {Liu}}]{sun2017substrate}%
  \BibitemOpen
  \bibfield  {author} {\bibinfo {author} {\bibfnamefont {Y.}~\bibnamefont
  {Sun}}, \bibinfo {author} {\bibfnamefont {R.}~\bibnamefont {Wang}},\ and\
  \bibinfo {author} {\bibfnamefont {K.}~\bibnamefont {Liu}},\ }\bibfield
  {title} {\bibinfo {title} {Substrate induced changes in atomically thin
  2-dimensional semiconductors: Fundamentals, engineering, and applications},\
  }\bibfield  {journal} {\bibinfo  {journal} {Applied Physics Reviews}\
  }\textbf {\bibinfo {volume} {4}},\ \href {https://doi.org/10.1063/1.4974072}
  {10.1063/1.4974072} (\bibinfo {year} {2017})\BibitemShut {NoStop}%
\bibitem [{\citenamefont {Lu}\ \emph {et~al.}(2014)\citenamefont {Lu},
  \citenamefont {Li}, \citenamefont {Watanabe}, \citenamefont {Taniguchi},\
  and\ \citenamefont {Andrei}}]{lu2014mos}%
  \BibitemOpen
  \bibfield  {author} {\bibinfo {author} {\bibfnamefont {C.-P.}\ \bibnamefont
  {Lu}}, \bibinfo {author} {\bibfnamefont {G.}~\bibnamefont {Li}}, \bibinfo
  {author} {\bibfnamefont {K.}~\bibnamefont {Watanabe}}, \bibinfo {author}
  {\bibfnamefont {T.}~\bibnamefont {Taniguchi}},\ and\ \bibinfo {author}
  {\bibfnamefont {E.~Y.}\ \bibnamefont {Andrei}},\ }\bibfield  {title}
  {\bibinfo {title} {Mos 2: choice substrate for accessing and tuning the
  electronic properties of graphene},\ }\href
  {https://doi.org/10.1103/PhysRevLett.113.156804} {\bibfield  {journal}
  {\bibinfo  {journal} {Physical review letters}\ }\textbf {\bibinfo {volume}
  {113}},\ \bibinfo {pages} {156804} (\bibinfo {year} {2014})}\BibitemShut
  {NoStop}%
\bibitem [{\citenamefont {Roy}\ \emph {et~al.}(2013)\citenamefont {Roy},
  \citenamefont {Padmanabhan}, \citenamefont {Goswami}, \citenamefont {Sai},
  \citenamefont {Ramalingam}, \citenamefont {Raghavan},\ and\ \citenamefont
  {Ghosh}}]{roy2013graphene}%
  \BibitemOpen
  \bibfield  {author} {\bibinfo {author} {\bibfnamefont {K.}~\bibnamefont
  {Roy}}, \bibinfo {author} {\bibfnamefont {M.}~\bibnamefont {Padmanabhan}},
  \bibinfo {author} {\bibfnamefont {S.}~\bibnamefont {Goswami}}, \bibinfo
  {author} {\bibfnamefont {T.~P.}\ \bibnamefont {Sai}}, \bibinfo {author}
  {\bibfnamefont {G.}~\bibnamefont {Ramalingam}}, \bibinfo {author}
  {\bibfnamefont {S.}~\bibnamefont {Raghavan}},\ and\ \bibinfo {author}
  {\bibfnamefont {A.}~\bibnamefont {Ghosh}},\ }\bibfield  {title} {\bibinfo
  {title} {Graphene--mos2 hybrid structures for multifunctional photoresponsive
  memory devices},\ }\href {https://doi.org/10.1038/nnano.2013.206} {\bibfield
  {journal} {\bibinfo  {journal} {Nature nanotechnology}\ }\textbf {\bibinfo
  {volume} {8}},\ \bibinfo {pages} {826} (\bibinfo {year} {2013})}\BibitemShut
  {NoStop}%
\bibitem [{\citenamefont {Wang}\ \emph {et~al.}(2015)\citenamefont {Wang},
  \citenamefont {Tad{\'e}},\ and\ \citenamefont {Shao}}]{wang2015research}%
  \BibitemOpen
  \bibfield  {author} {\bibinfo {author} {\bibfnamefont {W.}~\bibnamefont
  {Wang}}, \bibinfo {author} {\bibfnamefont {M.~O.}\ \bibnamefont {Tad{\'e}}},\
  and\ \bibinfo {author} {\bibfnamefont {Z.}~\bibnamefont {Shao}},\ }\bibfield
  {title} {\bibinfo {title} {Research progress of perovskite materials in
  photocatalysis-and photovoltaics-related energy conversion and environmental
  treatment},\ }\href {https://doi.org/10.1039/C5CS00113G} {\bibfield
  {journal} {\bibinfo  {journal} {Chemical Society Reviews}\ }\textbf {\bibinfo
  {volume} {44}},\ \bibinfo {pages} {5371} (\bibinfo {year}
  {2015})}\BibitemShut {NoStop}%
\bibitem [{\citenamefont {Rawat}\ \emph {et~al.}(2019)\citenamefont {Rawat},
  \citenamefont {Ahammed}, \citenamefont {Dimple}, \citenamefont {Jena},
  \citenamefont {Mohanta},\ and\ \citenamefont {De~Sarkar}}]{rawat2019solar}%
  \BibitemOpen
  \bibfield  {author} {\bibinfo {author} {\bibfnamefont {A.}~\bibnamefont
  {Rawat}}, \bibinfo {author} {\bibfnamefont {R.}~\bibnamefont {Ahammed}},
  \bibinfo {author} {\bibnamefont {Dimple}}, \bibinfo {author} {\bibfnamefont
  {N.}~\bibnamefont {Jena}}, \bibinfo {author} {\bibfnamefont {M.~K.}\
  \bibnamefont {Mohanta}},\ and\ \bibinfo {author} {\bibfnamefont
  {A.}~\bibnamefont {De~Sarkar}},\ }\bibfield  {title} {\bibinfo {title} {Solar
  energy harvesting in type ii van der waals heterostructures of semiconducting
  group iii monochalcogenide monolayers},\ }\href
  {https://doi.org/10.1021/acs.jpcc.9b03359} {\bibfield  {journal} {\bibinfo
  {journal} {The Journal of Physical Chemistry C}\ }\textbf {\bibinfo {volume}
  {123}},\ \bibinfo {pages} {12666} (\bibinfo {year} {2019})}\BibitemShut
  {NoStop}%
\bibitem [{\citenamefont {Mao}\ \emph {et~al.}(2023)\citenamefont {Mao},
  \citenamefont {Qin}, \citenamefont {Zhou}, \citenamefont {Zhang},\ and\
  \citenamefont {Yuan}}]{mao2023first}%
  \BibitemOpen
  \bibfield  {author} {\bibinfo {author} {\bibfnamefont {Y.}~\bibnamefont
  {Mao}}, \bibinfo {author} {\bibfnamefont {C.}~\bibnamefont {Qin}}, \bibinfo
  {author} {\bibfnamefont {X.}~\bibnamefont {Zhou}}, \bibinfo {author}
  {\bibfnamefont {Z.}~\bibnamefont {Zhang}},\ and\ \bibinfo {author}
  {\bibfnamefont {J.}~\bibnamefont {Yuan}},\ }\bibfield  {title} {\bibinfo
  {title} {First-principles study on gec/$\beta$-asp heterostructure with
  type-ii band alignment for photocatalytic water splitting},\ }\href
  {https://doi.org/10.1016/j.apsusc.2022.156298} {\bibfield  {journal}
  {\bibinfo  {journal} {Applied Surface Science}\ }\textbf {\bibinfo {volume}
  {617}},\ \bibinfo {pages} {156298} (\bibinfo {year} {2023})}\BibitemShut
  {NoStop}%
\bibitem [{\citenamefont {Li}\ \emph {et~al.}(2021)\citenamefont {Li},
  \citenamefont {Wang}, \citenamefont {Wang}, \citenamefont {Yang},
  \citenamefont {Zhao}, \citenamefont {Jia},\ and\ \citenamefont
  {Ke}}]{li2021two}%
  \BibitemOpen
  \bibfield  {author} {\bibinfo {author} {\bibfnamefont {X.-H.}\ \bibnamefont
  {Li}}, \bibinfo {author} {\bibfnamefont {B.-J.}\ \bibnamefont {Wang}},
  \bibinfo {author} {\bibfnamefont {G.-D.}\ \bibnamefont {Wang}}, \bibinfo
  {author} {\bibfnamefont {X.-F.}\ \bibnamefont {Yang}}, \bibinfo {author}
  {\bibfnamefont {R.-Q.}\ \bibnamefont {Zhao}}, \bibinfo {author}
  {\bibfnamefont {X.-T.}\ \bibnamefont {Jia}},\ and\ \bibinfo {author}
  {\bibfnamefont {S.-H.}\ \bibnamefont {Ke}},\ }\bibfield  {title} {\bibinfo
  {title} {A two-dimensional arsenene/gc 3 n 4 van der waals heterostructure: a
  highly efficient photocatalyst for water splitting},\ }\href
  {https://doi.org/10.1039/D1SE00313E} {\bibfield  {journal} {\bibinfo
  {journal} {Sustainable Energy \& Fuels}\ }\textbf {\bibinfo {volume} {5}},\
  \bibinfo {pages} {2249} (\bibinfo {year} {2021})}\BibitemShut {NoStop}%
\bibitem [{\citenamefont {He}\ \emph {et~al.}(2019)\citenamefont {He},
  \citenamefont {Zhang}, \citenamefont {Zhang},\ and\ \citenamefont
  {Li}}]{he2019type}%
  \BibitemOpen
  \bibfield  {author} {\bibinfo {author} {\bibfnamefont {C.}~\bibnamefont
  {He}}, \bibinfo {author} {\bibfnamefont {J.}~\bibnamefont {Zhang}}, \bibinfo
  {author} {\bibfnamefont {W.}~\bibnamefont {Zhang}},\ and\ \bibinfo {author}
  {\bibfnamefont {T.}~\bibnamefont {Li}},\ }\bibfield  {title} {\bibinfo
  {title} {Type-ii inse/g-c3n4 heterostructure as a high-efficiency oxygen
  evolution reaction catalyst for photoelectrochemical water splitting},\
  }\href {https://doi.org/10.1021/acs.jpclett.9b00909} {\bibfield  {journal}
  {\bibinfo  {journal} {The journal of physical chemistry letters}\ }\textbf
  {\bibinfo {volume} {10}},\ \bibinfo {pages} {3122} (\bibinfo {year}
  {2019})}\BibitemShut {NoStop}%
\bibitem [{\citenamefont {Wang}\ \emph {et~al.}(2013)\citenamefont {Wang},
  \citenamefont {Wang}, \citenamefont {Zhan}, \citenamefont {Wang},
  \citenamefont {Safdar},\ and\ \citenamefont {He}}]{wang2013visible}%
  \BibitemOpen
  \bibfield  {author} {\bibinfo {author} {\bibfnamefont {Y.}~\bibnamefont
  {Wang}}, \bibinfo {author} {\bibfnamefont {Q.}~\bibnamefont {Wang}}, \bibinfo
  {author} {\bibfnamefont {X.}~\bibnamefont {Zhan}}, \bibinfo {author}
  {\bibfnamefont {F.}~\bibnamefont {Wang}}, \bibinfo {author} {\bibfnamefont
  {M.}~\bibnamefont {Safdar}},\ and\ \bibinfo {author} {\bibfnamefont
  {J.}~\bibnamefont {He}},\ }\bibfield  {title} {\bibinfo {title} {Visible
  light driven type ii heterostructures and their enhanced photocatalysis
  properties: a review},\ }\href {https://doi.org/10.1039/C3NR01577G}
  {\bibfield  {journal} {\bibinfo  {journal} {Nanoscale}\ }\textbf {\bibinfo
  {volume} {5}},\ \bibinfo {pages} {8326} (\bibinfo {year} {2013})}\BibitemShut
  {NoStop}%
\bibitem [{\citenamefont {Priydarshi}\ \emph {et~al.}(2023)\citenamefont
  {Priydarshi}, \citenamefont {Arora}, \citenamefont {Chauhan}, \citenamefont
  {Agarwal},\ and\ \citenamefont {Bhowmick}}]{priydarshi2023versatility}%
  \BibitemOpen
  \bibfield  {author} {\bibinfo {author} {\bibfnamefont {A.}~\bibnamefont
  {Priydarshi}}, \bibinfo {author} {\bibfnamefont {A.}~\bibnamefont {Arora}},
  \bibinfo {author} {\bibfnamefont {Y.~S.}\ \bibnamefont {Chauhan}}, \bibinfo
  {author} {\bibfnamefont {A.}~\bibnamefont {Agarwal}},\ and\ \bibinfo {author}
  {\bibfnamefont {S.}~\bibnamefont {Bhowmick}},\ }\bibfield  {title} {\bibinfo
  {title} {Versatility of the sih--cdcl2 heterostructure: Piezoelectricity,
  photocatalysis, and transistor applications},\ }\href
  {https://doi.org/10.1021/acs.jpcc.3c04021} {\bibfield  {journal} {\bibinfo
  {journal} {The Journal of Physical Chemistry C}\ }\textbf {\bibinfo {volume}
  {127}},\ \bibinfo {pages} {21279} (\bibinfo {year} {2023})}\BibitemShut
  {NoStop}%
\bibitem [{\citenamefont {Kim}\ \emph {et~al.}(2015)\citenamefont {Kim},
  \citenamefont {Larentis}, \citenamefont {Fallahazad}, \citenamefont {Lee},
  \citenamefont {Xue}, \citenamefont {Dillen}, \citenamefont {Corbet},\ and\
  \citenamefont {Tutuc}}]{kim2015band}%
  \BibitemOpen
  \bibfield  {author} {\bibinfo {author} {\bibfnamefont {K.}~\bibnamefont
  {Kim}}, \bibinfo {author} {\bibfnamefont {S.}~\bibnamefont {Larentis}},
  \bibinfo {author} {\bibfnamefont {B.}~\bibnamefont {Fallahazad}}, \bibinfo
  {author} {\bibfnamefont {K.}~\bibnamefont {Lee}}, \bibinfo {author}
  {\bibfnamefont {J.}~\bibnamefont {Xue}}, \bibinfo {author} {\bibfnamefont
  {D.~C.}\ \bibnamefont {Dillen}}, \bibinfo {author} {\bibfnamefont {C.~M.}\
  \bibnamefont {Corbet}},\ and\ \bibinfo {author} {\bibfnamefont
  {E.}~\bibnamefont {Tutuc}},\ }\bibfield  {title} {\bibinfo {title} {Band
  alignment in wse2--graphene heterostructures},\ }\href
  {https://doi.org/10.1021/acsnano.5b01114} {\bibfield  {journal} {\bibinfo
  {journal} {ACS nano}\ }\textbf {\bibinfo {volume} {9}},\ \bibinfo {pages}
  {4527} (\bibinfo {year} {2015})}\BibitemShut {NoStop}%
\bibitem [{\citenamefont {Kang}\ \emph {et~al.}(2013)\citenamefont {Kang},
  \citenamefont {Tongay}, \citenamefont {Zhou}, \citenamefont {Li},\ and\
  \citenamefont {Wu}}]{kang2013band}%
  \BibitemOpen
  \bibfield  {author} {\bibinfo {author} {\bibfnamefont {J.}~\bibnamefont
  {Kang}}, \bibinfo {author} {\bibfnamefont {S.}~\bibnamefont {Tongay}},
  \bibinfo {author} {\bibfnamefont {J.}~\bibnamefont {Zhou}}, \bibinfo {author}
  {\bibfnamefont {J.}~\bibnamefont {Li}},\ and\ \bibinfo {author}
  {\bibfnamefont {J.}~\bibnamefont {Wu}},\ }\bibfield  {title} {\bibinfo
  {title} {Band offsets and heterostructures of two-dimensional
  semiconductors},\ }\bibfield  {journal} {\bibinfo  {journal} {Applied Physics
  Letters}\ }\textbf {\bibinfo {volume} {102}},\ \href
  {https://doi.org/10.1063/1.4774090} {10.1063/1.4774090} (\bibinfo {year}
  {2013})\BibitemShut {NoStop}%
\bibitem [{\citenamefont {Gong}\ \emph {et~al.}(2013)\citenamefont {Gong},
  \citenamefont {Zhang}, \citenamefont {Wang}, \citenamefont {Colombo},
  \citenamefont {Wallace},\ and\ \citenamefont {Cho}}]{gong2013band}%
  \BibitemOpen
  \bibfield  {author} {\bibinfo {author} {\bibfnamefont {C.}~\bibnamefont
  {Gong}}, \bibinfo {author} {\bibfnamefont {H.}~\bibnamefont {Zhang}},
  \bibinfo {author} {\bibfnamefont {W.}~\bibnamefont {Wang}}, \bibinfo {author}
  {\bibfnamefont {L.}~\bibnamefont {Colombo}}, \bibinfo {author} {\bibfnamefont
  {R.~M.}\ \bibnamefont {Wallace}},\ and\ \bibinfo {author} {\bibfnamefont
  {K.}~\bibnamefont {Cho}},\ }\bibfield  {title} {\bibinfo {title} {Band
  alignment of two-dimensional transition metal dichalcogenides: Application in
  tunnel field effect transistors},\ }\bibfield  {journal} {\bibinfo  {journal}
  {Applied Physics Letters}\ }\textbf {\bibinfo {volume} {103}},\ \href
  {https://doi.org/10.1063/1.4817409} {10.1063/1.4817409} (\bibinfo {year}
  {2013})\BibitemShut {NoStop}%
\bibitem [{\citenamefont {Rahman}\ \emph {et~al.}(2016)\citenamefont {Rahman},
  \citenamefont {Kwong}, \citenamefont {Davey},\ and\ \citenamefont
  {Qiao}}]{rahman20162d}%
  \BibitemOpen
  \bibfield  {author} {\bibinfo {author} {\bibfnamefont {M.~Z.}\ \bibnamefont
  {Rahman}}, \bibinfo {author} {\bibfnamefont {C.~W.}\ \bibnamefont {Kwong}},
  \bibinfo {author} {\bibfnamefont {K.}~\bibnamefont {Davey}},\ and\ \bibinfo
  {author} {\bibfnamefont {S.~Z.}\ \bibnamefont {Qiao}},\ }\bibfield  {title}
  {\bibinfo {title} {2d phosphorene as a water splitting photocatalyst:
  fundamentals to applications},\ }\href {https://doi.org/10.1039/C5EE03732H}
  {\bibfield  {journal} {\bibinfo  {journal} {Energy \& Environmental Science}\
  }\textbf {\bibinfo {volume} {9}},\ \bibinfo {pages} {709} (\bibinfo {year}
  {2016})}\BibitemShut {NoStop}%
\bibitem [{\citenamefont {Guo}\ \emph {et~al.}(2016)\citenamefont {Guo},
  \citenamefont {Zhou}, \citenamefont {Zhu},\ and\ \citenamefont
  {Sun}}]{guo2016mxene}%
  \BibitemOpen
  \bibfield  {author} {\bibinfo {author} {\bibfnamefont {Z.}~\bibnamefont
  {Guo}}, \bibinfo {author} {\bibfnamefont {J.}~\bibnamefont {Zhou}}, \bibinfo
  {author} {\bibfnamefont {L.}~\bibnamefont {Zhu}},\ and\ \bibinfo {author}
  {\bibfnamefont {Z.}~\bibnamefont {Sun}},\ }\bibfield  {title} {\bibinfo
  {title} {Mxene: a promising photocatalyst for water splitting},\ }\href
  {https://doi.org/10.1039/C6TA04414J} {\bibfield  {journal} {\bibinfo
  {journal} {Journal of Materials Chemistry A}\ }\textbf {\bibinfo {volume}
  {4}},\ \bibinfo {pages} {11446} (\bibinfo {year} {2016})}\BibitemShut
  {NoStop}%
\bibitem [{\citenamefont {Zhan}\ \emph {et~al.}(2019)\citenamefont {Zhan},
  \citenamefont {Sun}, \citenamefont {Xie}, \citenamefont {Jiang},\ and\
  \citenamefont {Kent}}]{zhan2019computational}%
  \BibitemOpen
  \bibfield  {author} {\bibinfo {author} {\bibfnamefont {C.}~\bibnamefont
  {Zhan}}, \bibinfo {author} {\bibfnamefont {W.}~\bibnamefont {Sun}}, \bibinfo
  {author} {\bibfnamefont {Y.}~\bibnamefont {Xie}}, \bibinfo {author}
  {\bibfnamefont {D.-e.}\ \bibnamefont {Jiang}},\ and\ \bibinfo {author}
  {\bibfnamefont {P.~R.}\ \bibnamefont {Kent}},\ }\bibfield  {title} {\bibinfo
  {title} {Computational discovery and design of mxenes for energy
  applications: status, successes, and opportunities},\ }\href
  {https://doi.org/10.1021/acsami.9b00439} {\bibfield  {journal} {\bibinfo
  {journal} {ACS applied materials \& interfaces}\ }\textbf {\bibinfo {volume}
  {11}},\ \bibinfo {pages} {24885} (\bibinfo {year} {2019})}\BibitemShut
  {NoStop}%
\bibitem [{\citenamefont {Colibaba}\ \emph {et~al.}(2019)\citenamefont
  {Colibaba}, \citenamefont {K\"orbel}, \citenamefont {Motta}, \citenamefont
  {El-Mellouhi},\ and\ \citenamefont {Sanvito}}]{PhysRevMaterials.3.124002}%
  \BibitemOpen
  \bibfield  {author} {\bibinfo {author} {\bibfnamefont {S.~A.}\ \bibnamefont
  {Colibaba}}, \bibinfo {author} {\bibfnamefont {S.}~\bibnamefont {K\"orbel}},
  \bibinfo {author} {\bibfnamefont {C.}~\bibnamefont {Motta}}, \bibinfo
  {author} {\bibfnamefont {F.}~\bibnamefont {El-Mellouhi}},\ and\ \bibinfo
  {author} {\bibfnamefont {S.}~\bibnamefont {Sanvito}},\ }\bibfield  {title}
  {\bibinfo {title} {Interlayer dielectric function of a type-ii van der waals
  semiconductor: The ${\mathrm{hfs}}_{2}/{\mathrm{pts}}_{2}$ heterobilayer},\
  }\href {https://doi.org/10.1103/PhysRevMaterials.3.124002} {\bibfield
  {journal} {\bibinfo  {journal} {Phys. Rev. Mater.}\ }\textbf {\bibinfo
  {volume} {3}},\ \bibinfo {pages} {124002} (\bibinfo {year}
  {2019})}\BibitemShut {NoStop}%
\bibitem [{\citenamefont {Gillen}\ and\ \citenamefont
  {Maultzsch}(2018)}]{PhysRevB.97.165306}%
  \BibitemOpen
  \bibfield  {author} {\bibinfo {author} {\bibfnamefont {R.}~\bibnamefont
  {Gillen}}\ and\ \bibinfo {author} {\bibfnamefont {J.}~\bibnamefont
  {Maultzsch}},\ }\bibfield  {title} {\bibinfo {title} {Interlayer excitons in
  ${\mathrm{mose}}_{2}\text{/}{\mathrm{wse}}_{2}$ heterostructures from first
  principles},\ }\href {https://doi.org/10.1103/PhysRevB.97.165306} {\bibfield
  {journal} {\bibinfo  {journal} {Phys. Rev. B}\ }\textbf {\bibinfo {volume}
  {97}},\ \bibinfo {pages} {165306} (\bibinfo {year} {2018})}\BibitemShut
  {NoStop}%
\bibitem [{\citenamefont {He}\ \emph {et~al.}(2020)\citenamefont {He},
  \citenamefont {Ma}, \citenamefont {Lei}, \citenamefont {Peng}, \citenamefont
  {Huang},\ and\ \citenamefont {Dai}}]{he2020tl2o}%
  \BibitemOpen
  \bibfield  {author} {\bibinfo {author} {\bibfnamefont {Z.}~\bibnamefont
  {He}}, \bibinfo {author} {\bibfnamefont {Y.}~\bibnamefont {Ma}}, \bibinfo
  {author} {\bibfnamefont {C.}~\bibnamefont {Lei}}, \bibinfo {author}
  {\bibfnamefont {R.}~\bibnamefont {Peng}}, \bibinfo {author} {\bibfnamefont
  {B.}~\bibnamefont {Huang}},\ and\ \bibinfo {author} {\bibfnamefont
  {Y.}~\bibnamefont {Dai}},\ }\bibfield  {title} {\bibinfo {title} {Tl2o/wte2
  van der waals heterostructure with tunable multiple band alignments},\
  }\bibfield  {journal} {\bibinfo  {journal} {The Journal of Chemical Physics}\
  }\textbf {\bibinfo {volume} {152}},\ \href
  {https://doi.org/10.1063/1.5141053} {10.1063/1.5141053} (\bibinfo {year}
  {2020})\BibitemShut {NoStop}%
\bibitem [{\citenamefont {Mounet}\ \emph {et~al.}(2018)\citenamefont {Mounet},
  \citenamefont {Gibertini}, \citenamefont {Schwaller}, \citenamefont {Campi},
  \citenamefont {Merkys}, \citenamefont {Marrazzo}, \citenamefont {Sohier},
  \citenamefont {Castelli}, \citenamefont {Cepellotti}, \citenamefont {Pizzi}
  \emph {et~al.}}]{mounet2018two}%
  \BibitemOpen
  \bibfield  {author} {\bibinfo {author} {\bibfnamefont {N.}~\bibnamefont
  {Mounet}}, \bibinfo {author} {\bibfnamefont {M.}~\bibnamefont {Gibertini}},
  \bibinfo {author} {\bibfnamefont {P.}~\bibnamefont {Schwaller}}, \bibinfo
  {author} {\bibfnamefont {D.}~\bibnamefont {Campi}}, \bibinfo {author}
  {\bibfnamefont {A.}~\bibnamefont {Merkys}}, \bibinfo {author} {\bibfnamefont
  {A.}~\bibnamefont {Marrazzo}}, \bibinfo {author} {\bibfnamefont
  {T.}~\bibnamefont {Sohier}}, \bibinfo {author} {\bibfnamefont {I.~E.}\
  \bibnamefont {Castelli}}, \bibinfo {author} {\bibfnamefont {A.}~\bibnamefont
  {Cepellotti}}, \bibinfo {author} {\bibfnamefont {G.}~\bibnamefont {Pizzi}},
  \emph {et~al.},\ }\bibfield  {title} {\bibinfo {title} {Two-dimensional
  materials from high-throughput computational exfoliation of experimentally
  known compounds},\ }\href {https://doi.org/10.1038/s41565-017-0035-5}
  {\bibfield  {journal} {\bibinfo  {journal} {Nature nanotechnology}\ }\textbf
  {\bibinfo {volume} {13}},\ \bibinfo {pages} {246} (\bibinfo {year}
  {2018})}\BibitemShut {NoStop}%
\bibitem [{\citenamefont {Giannozzi}\ \emph {et~al.}(2009)\citenamefont
  {Giannozzi}, \citenamefont {Baroni}, \citenamefont {Bonini}, \citenamefont
  {Calandra}, \citenamefont {Car}, \citenamefont {Cavazzoni}, \citenamefont
  {Ceresoli}, \citenamefont {Chiarotti}, \citenamefont {Cococcioni},
  \citenamefont {Dabo} \emph {et~al.}}]{giannozzi2009quantum}%
  \BibitemOpen
  \bibfield  {author} {\bibinfo {author} {\bibfnamefont {P.}~\bibnamefont
  {Giannozzi}}, \bibinfo {author} {\bibfnamefont {S.}~\bibnamefont {Baroni}},
  \bibinfo {author} {\bibfnamefont {N.}~\bibnamefont {Bonini}}, \bibinfo
  {author} {\bibfnamefont {M.}~\bibnamefont {Calandra}}, \bibinfo {author}
  {\bibfnamefont {R.}~\bibnamefont {Car}}, \bibinfo {author} {\bibfnamefont
  {C.}~\bibnamefont {Cavazzoni}}, \bibinfo {author} {\bibfnamefont
  {D.}~\bibnamefont {Ceresoli}}, \bibinfo {author} {\bibfnamefont {G.~L.}\
  \bibnamefont {Chiarotti}}, \bibinfo {author} {\bibfnamefont {M.}~\bibnamefont
  {Cococcioni}}, \bibinfo {author} {\bibfnamefont {I.}~\bibnamefont {Dabo}},
  \emph {et~al.},\ }\bibfield  {title} {\bibinfo {title} {Quantum espresso: a
  modular and open-source software project for quantum simulations of
  materials},\ }\href {https://doi.org/10.1088/0953-8984/21/39/395502}
  {\bibfield  {journal} {\bibinfo  {journal} {Journal of physics: Condensed
  matter}\ }\textbf {\bibinfo {volume} {21}},\ \bibinfo {pages} {395502}
  (\bibinfo {year} {2009})}\BibitemShut {NoStop}%
\bibitem [{\citenamefont {Perdew}\ \emph {et~al.}(1996)\citenamefont {Perdew},
  \citenamefont {Burke},\ and\ \citenamefont
  {Ernzerhof}}]{perdew1996generalized}%
  \BibitemOpen
  \bibfield  {author} {\bibinfo {author} {\bibfnamefont {J.~P.}\ \bibnamefont
  {Perdew}}, \bibinfo {author} {\bibfnamefont {K.}~\bibnamefont {Burke}},\ and\
  \bibinfo {author} {\bibfnamefont {M.}~\bibnamefont {Ernzerhof}},\ }\bibfield
  {title} {\bibinfo {title} {Generalized gradient approximation made simple},\
  }\href {https://doi.org/10.1103/PhysRevLett.77.3865} {\bibfield  {journal}
  {\bibinfo  {journal} {Physical review letters}\ }\textbf {\bibinfo {volume}
  {77}},\ \bibinfo {pages} {3865} (\bibinfo {year} {1996})}\BibitemShut
  {NoStop}%
\bibitem [{\citenamefont {Ren}\ \emph {et~al.}(2019)\citenamefont {Ren},
  \citenamefont {Ren}, \citenamefont {Luo}, \citenamefont {Xu}, \citenamefont
  {Yu}, \citenamefont {Tang},\ and\ \citenamefont {Sun}}]{ren2019using}%
  \BibitemOpen
  \bibfield  {author} {\bibinfo {author} {\bibfnamefont {K.}~\bibnamefont
  {Ren}}, \bibinfo {author} {\bibfnamefont {C.}~\bibnamefont {Ren}}, \bibinfo
  {author} {\bibfnamefont {Y.}~\bibnamefont {Luo}}, \bibinfo {author}
  {\bibfnamefont {Y.}~\bibnamefont {Xu}}, \bibinfo {author} {\bibfnamefont
  {J.}~\bibnamefont {Yu}}, \bibinfo {author} {\bibfnamefont {W.}~\bibnamefont
  {Tang}},\ and\ \bibinfo {author} {\bibfnamefont {M.}~\bibnamefont {Sun}},\
  }\bibfield  {title} {\bibinfo {title} {Using van der waals heterostructures
  based on two-dimensional blue phosphorus and xc (x= ge, si) for
  water-splitting photocatalysis: a first-principles study},\ }\href
  {https://doi.org/10.1039/C8CP07680D} {\bibfield  {journal} {\bibinfo
  {journal} {Physical Chemistry Chemical Physics}\ }\textbf {\bibinfo {volume}
  {21}},\ \bibinfo {pages} {9949} (\bibinfo {year} {2019})}\BibitemShut
  {NoStop}%
\bibitem [{\citenamefont {Pack}\ and\ \citenamefont
  {Monkhorst}(1977)}]{pack1977special}%
  \BibitemOpen
  \bibfield  {author} {\bibinfo {author} {\bibfnamefont {J.~D.}\ \bibnamefont
  {Pack}}\ and\ \bibinfo {author} {\bibfnamefont {H.~J.}\ \bibnamefont
  {Monkhorst}},\ }\bibfield  {title} {\bibinfo {title} {" special points for
  brillouin-zone integrations"—a reply},\ }\href
  {https://doi.org/10.1103/PhysRevB.16.1748} {\bibfield  {journal} {\bibinfo
  {journal} {Physical Review B}\ }\textbf {\bibinfo {volume} {16}},\ \bibinfo
  {pages} {1748} (\bibinfo {year} {1977})}\BibitemShut {NoStop}%
\bibitem [{\citenamefont {Grimme}(2006)}]{grimme2006semiempirical}%
  \BibitemOpen
  \bibfield  {author} {\bibinfo {author} {\bibfnamefont {S.}~\bibnamefont
  {Grimme}},\ }\bibfield  {title} {\bibinfo {title} {Semiempirical gga-type
  density functional constructed with a long-range dispersion correction},\
  }\href {https://doi.org/10.1002/jcc.20495} {\bibfield  {journal} {\bibinfo
  {journal} {Journal of computational chemistry}\ }\textbf {\bibinfo {volume}
  {27}},\ \bibinfo {pages} {1787} (\bibinfo {year} {2006})}\BibitemShut
  {NoStop}%
\bibitem [{\citenamefont {Zhang}\ \emph {et~al.}(2017)\citenamefont {Zhang},
  \citenamefont {Chen}, \citenamefont {Duan}, \citenamefont {Zang},
  \citenamefont {Luo},\ and\ \citenamefont {Duan}}]{zhang2017robust}%
  \BibitemOpen
  \bibfield  {author} {\bibinfo {author} {\bibfnamefont {Z.}~\bibnamefont
  {Zhang}}, \bibinfo {author} {\bibfnamefont {P.}~\bibnamefont {Chen}},
  \bibinfo {author} {\bibfnamefont {X.}~\bibnamefont {Duan}}, \bibinfo {author}
  {\bibfnamefont {K.}~\bibnamefont {Zang}}, \bibinfo {author} {\bibfnamefont
  {J.}~\bibnamefont {Luo}},\ and\ \bibinfo {author} {\bibfnamefont
  {X.}~\bibnamefont {Duan}},\ }\bibfield  {title} {\bibinfo {title} {Robust
  epitaxial growth of two-dimensional heterostructures, multiheterostructures,
  and superlattices},\ }\href {https://doi.org/10.1126/science.aan6814}
  {\bibfield  {journal} {\bibinfo  {journal} {Science}\ }\textbf {\bibinfo
  {volume} {357}},\ \bibinfo {pages} {788} (\bibinfo {year}
  {2017})}\BibitemShut {NoStop}%
\bibitem [{\citenamefont {Gajdo\ifmmode~\check{s}\else \v{s}\fi{}}\ \emph
  {et~al.}(2006)\citenamefont {Gajdo\ifmmode~\check{s}\else \v{s}\fi{}},
  \citenamefont {Hummer}, \citenamefont {Kresse}, \citenamefont
  {Furthm\"uller},\ and\ \citenamefont {Bechstedt}}]{PhysRevB.73.045112}%
  \BibitemOpen
  \bibfield  {author} {\bibinfo {author} {\bibfnamefont {M.}~\bibnamefont
  {Gajdo\ifmmode~\check{s}\else \v{s}\fi{}}}, \bibinfo {author} {\bibfnamefont
  {K.}~\bibnamefont {Hummer}}, \bibinfo {author} {\bibfnamefont
  {G.}~\bibnamefont {Kresse}}, \bibinfo {author} {\bibfnamefont
  {J.}~\bibnamefont {Furthm\"uller}},\ and\ \bibinfo {author} {\bibfnamefont
  {F.}~\bibnamefont {Bechstedt}},\ }\bibfield  {title} {\bibinfo {title}
  {Linear optical properties in the projector-augmented wave methodology},\
  }\href {https://doi.org/10.1103/PhysRevB.73.045112} {\bibfield  {journal}
  {\bibinfo  {journal} {Phys. Rev. B}\ }\textbf {\bibinfo {volume} {73}},\
  \bibinfo {pages} {045112} (\bibinfo {year} {2006})}\BibitemShut {NoStop}%
\bibitem [{\citenamefont {Bernardi}\ \emph
  {et~al.}(2013{\natexlab{b}})\citenamefont {Bernardi}, \citenamefont
  {Palummo},\ and\ \citenamefont {Grossman}}]{1bernardi2013extraordinary}%
  \BibitemOpen
  \bibfield  {author} {\bibinfo {author} {\bibfnamefont {M.}~\bibnamefont
  {Bernardi}}, \bibinfo {author} {\bibfnamefont {M.}~\bibnamefont {Palummo}},\
  and\ \bibinfo {author} {\bibfnamefont {J.~C.}\ \bibnamefont {Grossman}},\
  }\bibfield  {title} {\bibinfo {title} {Extraordinary sunlight absorption and
  one nanometer thick photovoltaics using two-dimensional monolayer
  materials},\ }\href {https://doi.org/10.1021/nl401544y} {\bibfield  {journal}
  {\bibinfo  {journal} {Nano letters}\ }\textbf {\bibinfo {volume} {13}},\
  \bibinfo {pages} {3664} (\bibinfo {year} {2013}{\natexlab{b}})}\BibitemShut
  {NoStop}%
\bibitem [{\citenamefont {Furchi}\ \emph {et~al.}(2014)\citenamefont {Furchi},
  \citenamefont {Pospischil}, \citenamefont {Libisch}, \citenamefont
  {Burgdorfer},\ and\ \citenamefont {Mueller}}]{2furchi2014photovoltaic}%
  \BibitemOpen
  \bibfield  {author} {\bibinfo {author} {\bibfnamefont {M.~M.}\ \bibnamefont
  {Furchi}}, \bibinfo {author} {\bibfnamefont {A.}~\bibnamefont {Pospischil}},
  \bibinfo {author} {\bibfnamefont {F.}~\bibnamefont {Libisch}}, \bibinfo
  {author} {\bibfnamefont {J.}~\bibnamefont {Burgdorfer}},\ and\ \bibinfo
  {author} {\bibfnamefont {T.}~\bibnamefont {Mueller}},\ }\bibfield  {title}
  {\bibinfo {title} {Photovoltaic effect in an electrically tunable van der
  waals heterojunction},\ }\href {https://doi.org/10.1021/nl501962c} {\bibfield
   {journal} {\bibinfo  {journal} {Nano letters}\ }\textbf {\bibinfo {volume}
  {14}},\ \bibinfo {pages} {4785} (\bibinfo {year} {2014})}\BibitemShut
  {NoStop}%
\bibitem [{\citenamefont {Aparicio-Huacarpuma}\ \emph
  {et~al.}(2025)\citenamefont {Aparicio-Huacarpuma}, \citenamefont {Marinho},
  \citenamefont {Giozza}, \citenamefont {Silva}, \citenamefont {Kenfack-Sadem},
  \citenamefont {Dias},\ and\ \citenamefont {Ribeiro}}]{3aparicio20252d}%
  \BibitemOpen
  \bibfield  {author} {\bibinfo {author} {\bibfnamefont {B.~D.}\ \bibnamefont
  {Aparicio-Huacarpuma}}, \bibinfo {author} {\bibfnamefont {E.}~\bibnamefont
  {Marinho}}, \bibinfo {author} {\bibfnamefont {W.~F.}\ \bibnamefont {Giozza}},
  \bibinfo {author} {\bibfnamefont {A.~M.}\ \bibnamefont {Silva}}, \bibinfo
  {author} {\bibfnamefont {C.}~\bibnamefont {Kenfack-Sadem}}, \bibinfo {author}
  {\bibfnamefont {A.~C.}\ \bibnamefont {Dias}},\ and\ \bibinfo {author}
  {\bibfnamefont {L.~A.}\ \bibnamefont {Ribeiro}},\ }\bibfield  {title}
  {\bibinfo {title} {2d janus snses monolayers for solar energy conversion:
  insights from dft and excitonic analysis},\ }\bibfield  {journal} {\bibinfo
  {journal} {Nanoscale}\ }\href {https://doi.org/10.1039/D5NR01745A}
  {10.1039/D5NR01745A} (\bibinfo {year} {2025})\BibitemShut {NoStop}%
\bibitem [{\citenamefont {Peng}\ \emph {et~al.}(2016)\citenamefont {Peng},
  \citenamefont {Wang}, \citenamefont {Sa}, \citenamefont {Wu},\ and\
  \citenamefont {Sun}}]{4peng2016electronic}%
  \BibitemOpen
  \bibfield  {author} {\bibinfo {author} {\bibfnamefont {Q.}~\bibnamefont
  {Peng}}, \bibinfo {author} {\bibfnamefont {Z.}~\bibnamefont {Wang}}, \bibinfo
  {author} {\bibfnamefont {B.}~\bibnamefont {Sa}}, \bibinfo {author}
  {\bibfnamefont {B.}~\bibnamefont {Wu}},\ and\ \bibinfo {author}
  {\bibfnamefont {Z.}~\bibnamefont {Sun}},\ }\bibfield  {title} {\bibinfo
  {title} {Electronic structures and enhanced optical properties of blue
  phosphorene/transition metal dichalcogenides van der waals
  heterostructures},\ }\href
  {https://doi.org/https://doi.org/10.1038/srep31994} {\bibfield  {journal}
  {\bibinfo  {journal} {Scientific reports}\ }\textbf {\bibinfo {volume} {6}},\
  \bibinfo {pages} {31994} (\bibinfo {year} {2016})}\BibitemShut {NoStop}%
\bibitem [{\citenamefont {N{\o}rskov}\ \emph {et~al.}(2004)\citenamefont
  {N{\o}rskov}, \citenamefont {Rossmeisl}, \citenamefont {Logadottir},
  \citenamefont {Lindqvist}, \citenamefont {Kitchin}, \citenamefont
  {Bligaard},\ and\ \citenamefont {Jonsson}}]{norskov2004origin}%
  \BibitemOpen
  \bibfield  {author} {\bibinfo {author} {\bibfnamefont {J.~K.}\ \bibnamefont
  {N{\o}rskov}}, \bibinfo {author} {\bibfnamefont {J.}~\bibnamefont
  {Rossmeisl}}, \bibinfo {author} {\bibfnamefont {A.}~\bibnamefont
  {Logadottir}}, \bibinfo {author} {\bibfnamefont {L.}~\bibnamefont
  {Lindqvist}}, \bibinfo {author} {\bibfnamefont {J.~R.}\ \bibnamefont
  {Kitchin}}, \bibinfo {author} {\bibfnamefont {T.}~\bibnamefont {Bligaard}},\
  and\ \bibinfo {author} {\bibfnamefont {H.}~\bibnamefont {Jonsson}},\
  }\bibfield  {title} {\bibinfo {title} {Origin of the overpotential for oxygen
  reduction at a fuel-cell cathode},\ }\href
  {https://doi.org/10.1021/jp047349j} {\bibfield  {journal} {\bibinfo
  {journal} {The Journal of Physical Chemistry B}\ }\textbf {\bibinfo {volume}
  {108}},\ \bibinfo {pages} {17886} (\bibinfo {year} {2004})}\BibitemShut
  {NoStop}%
\bibitem [{\citenamefont {Vald{\'e}s}\ \emph {et~al.}(2008)\citenamefont
  {Vald{\'e}s}, \citenamefont {Qu}, \citenamefont {Kroes}, \citenamefont
  {Rossmeisl},\ and\ \citenamefont {N{\o}rskov}}]{valdes2008oxidation}%
  \BibitemOpen
  \bibfield  {author} {\bibinfo {author} {\bibfnamefont {{\'A}.}~\bibnamefont
  {Vald{\'e}s}}, \bibinfo {author} {\bibfnamefont {Z.-W.}\ \bibnamefont {Qu}},
  \bibinfo {author} {\bibfnamefont {G.-J.}\ \bibnamefont {Kroes}}, \bibinfo
  {author} {\bibfnamefont {J.}~\bibnamefont {Rossmeisl}},\ and\ \bibinfo
  {author} {\bibfnamefont {J.~K.}\ \bibnamefont {N{\o}rskov}},\ }\bibfield
  {title} {\bibinfo {title} {Oxidation and photo-oxidation of water on tio2
  surface},\ }\href {https://doi.org/10.1021/jp711929d} {\bibfield  {journal}
  {\bibinfo  {journal} {The Journal of Physical Chemistry C}\ }\textbf
  {\bibinfo {volume} {112}},\ \bibinfo {pages} {9872} (\bibinfo {year}
  {2008})}\BibitemShut {NoStop}%
\bibitem [{\citenamefont {Zhou}\ \emph {et~al.}(2016)\citenamefont {Zhou},
  \citenamefont {Chen}, \citenamefont {Cui}, \citenamefont {Zeng},
  \citenamefont {Lin}, \citenamefont {Kaxiras},\ and\ \citenamefont
  {Zhang}}]{nanolett.6b02052}%
  \BibitemOpen
  \bibfield  {author} {\bibinfo {author} {\bibfnamefont {Y.}~\bibnamefont
  {Zhou}}, \bibinfo {author} {\bibfnamefont {W.}~\bibnamefont {Chen}}, \bibinfo
  {author} {\bibfnamefont {P.}~\bibnamefont {Cui}}, \bibinfo {author}
  {\bibfnamefont {J.}~\bibnamefont {Zeng}}, \bibinfo {author} {\bibfnamefont
  {Z.}~\bibnamefont {Lin}}, \bibinfo {author} {\bibfnamefont {E.}~\bibnamefont
  {Kaxiras}},\ and\ \bibinfo {author} {\bibfnamefont {Z.}~\bibnamefont
  {Zhang}},\ }\bibfield  {title} {\bibinfo {title} {Enhancing the hydrogen
  activation reactivity of nonprecious metal substrates via confined catalysis
  underneath graphene},\ }\href {https://doi.org/10.1021/acs.nanolett.6b02052}
  {\bibfield  {journal} {\bibinfo  {journal} {Nano Letters}\ }\textbf {\bibinfo
  {volume} {16}},\ \bibinfo {pages} {6058} (\bibinfo {year}
  {2016})}\BibitemShut {NoStop}%
\bibitem [{\citenamefont {Nørskov}\ \emph {et~al.}(2005)\citenamefont
  {Nørskov}, \citenamefont {Bligaard}, \citenamefont {Logadottir},
  \citenamefont {Kitchin}, \citenamefont {Chen}, \citenamefont {Pandelov},\
  and\ \citenamefont {Stimming}}]{Nørskov_2005}%
  \BibitemOpen
  \bibfield  {author} {\bibinfo {author} {\bibfnamefont {J.~K.}\ \bibnamefont
  {Nørskov}}, \bibinfo {author} {\bibfnamefont {T.}~\bibnamefont {Bligaard}},
  \bibinfo {author} {\bibfnamefont {A.}~\bibnamefont {Logadottir}}, \bibinfo
  {author} {\bibfnamefont {J.~R.}\ \bibnamefont {Kitchin}}, \bibinfo {author}
  {\bibfnamefont {J.~G.}\ \bibnamefont {Chen}}, \bibinfo {author}
  {\bibfnamefont {S.}~\bibnamefont {Pandelov}},\ and\ \bibinfo {author}
  {\bibfnamefont {U.}~\bibnamefont {Stimming}},\ }\bibfield  {title} {\bibinfo
  {title} {Trends in the exchange current for hydrogen evolution},\ }\href
  {https://doi.org/10.1149/1.1856988} {\bibfield  {journal} {\bibinfo
  {journal} {Journal of The Electrochemical Society}\ }\textbf {\bibinfo
  {volume} {152}},\ \bibinfo {pages} {J23} (\bibinfo {year}
  {2005})}\BibitemShut {NoStop}%
\bibitem [{\citenamefont {Sahoo}\ \emph {et~al.}(2022)\citenamefont {Sahoo},
  \citenamefont {Ray},\ and\ \citenamefont {Singh}}]{acsomega.1c06730}%
  \BibitemOpen
  \bibfield  {author} {\bibinfo {author} {\bibfnamefont {M.~R.}\ \bibnamefont
  {Sahoo}}, \bibinfo {author} {\bibfnamefont {A.}~\bibnamefont {Ray}},\ and\
  \bibinfo {author} {\bibfnamefont {N.}~\bibnamefont {Singh}},\ }\bibfield
  {title} {\bibinfo {title} {Theoretical insights into the hydrogen evolution
  reaction on vge2n4 and nbge2n4 monolayers},\ }\href
  {https://doi.org/10.1021/acsomega.1c06730} {\bibfield  {journal} {\bibinfo
  {journal} {ACS Omega}\ }\textbf {\bibinfo {volume} {7}},\ \bibinfo {pages}
  {7837} (\bibinfo {year} {2022})}\BibitemShut {NoStop}%
\bibitem [{\citenamefont {Di~Liberto}\ \emph {et~al.}(2022)\citenamefont
  {Di~Liberto}, \citenamefont {Cipriano},\ and\ \citenamefont
  {Pacchioni}}]{di2022universal}%
  \BibitemOpen
  \bibfield  {author} {\bibinfo {author} {\bibfnamefont {G.}~\bibnamefont
  {Di~Liberto}}, \bibinfo {author} {\bibfnamefont {L.~A.}\ \bibnamefont
  {Cipriano}},\ and\ \bibinfo {author} {\bibfnamefont {G.}~\bibnamefont
  {Pacchioni}},\ }\bibfield  {title} {\bibinfo {title} {Universal principles
  for the rational design of single atom electrocatalysts? handle with care},\
  }\href {https://doi.org/10.1021/acscatal.2c01011} {\bibfield  {journal}
  {\bibinfo  {journal} {ACS Catalysis}\ }\textbf {\bibinfo {volume} {12}},\
  \bibinfo {pages} {5846} (\bibinfo {year} {2022})}\BibitemShut {NoStop}%
\bibitem [{\citenamefont {Qu}\ \emph {et~al.}(2017)\citenamefont {Qu},
  \citenamefont {Shao}, \citenamefont {Shao}, \citenamefont {Yang},
  \citenamefont {Xu}, \citenamefont {Kwok}, \citenamefont {Shi}, \citenamefont
  {Lu},\ and\ \citenamefont {Pan}}]{qu2017ultra}%
  \BibitemOpen
  \bibfield  {author} {\bibinfo {author} {\bibfnamefont {Y.}~\bibnamefont
  {Qu}}, \bibinfo {author} {\bibfnamefont {M.}~\bibnamefont {Shao}}, \bibinfo
  {author} {\bibfnamefont {Y.}~\bibnamefont {Shao}}, \bibinfo {author}
  {\bibfnamefont {M.}~\bibnamefont {Yang}}, \bibinfo {author} {\bibfnamefont
  {J.}~\bibnamefont {Xu}}, \bibinfo {author} {\bibfnamefont {C.~T.}\
  \bibnamefont {Kwok}}, \bibinfo {author} {\bibfnamefont {X.}~\bibnamefont
  {Shi}}, \bibinfo {author} {\bibfnamefont {Z.}~\bibnamefont {Lu}},\ and\
  \bibinfo {author} {\bibfnamefont {H.}~\bibnamefont {Pan}},\ }\bibfield
  {title} {\bibinfo {title} {Ultra-high electrocatalytic activity of vs 2
  nanoflowers for efficient hydrogen evolution reaction},\ }\href
  {https://doi.org/10.1039/C7TA03172F} {\bibfield  {journal} {\bibinfo
  {journal} {Journal of Materials Chemistry A}\ }\textbf {\bibinfo {volume}
  {5}},\ \bibinfo {pages} {15080} (\bibinfo {year} {2017})}\BibitemShut
  {NoStop}%
\end{thebibliography}%
\end{document}